\documentclass[11pt]{article}
\pdfoutput=1
\usepackage{jcapmod}

\usepackage{booktabs}
\usepackage{amsmath,amssymb,amsbsy,amstext, amsthm, simplewick}
\usepackage{hyperref}
\usepackage{graphicx}
\usepackage{enumitem}
\usepackage{amsfonts}
\usepackage{amssymb}
\usepackage[small]{caption}
\usepackage{upgreek}
\usepackage[svgnames,dvipsnames,x11names,table]{xcolor}
\usepackage{multirow} 
\usepackage{geometry}
\usepackage[hang,flushmargin]{footmisc} 
\usepackage[utf8]{inputenc} 

\usepackage{colortbl}
\definecolor{lightgreen}{cmyk}{0.2, 0, 0.2, 0.2}
\definecolor{lightgray}{cmyk}{0.1,0.2,0,0.1}
\definecolor{lightgray2}{cmyk}{0.1,0.1,0,0.1}

\makeatletter
\newlength{\apb@width}
\newcommand{\autoparbox}[2][c]{\settowidth{\apb@width}{#2}\parbox[#1]{\apb@width}{#2}}

\makeatother

\definecolor{lightgray}{gray}{0.9}

\usepackage[framemethod=default]{mdframed}
\newmdenv[skipabove=7pt,
skipbelow=7pt,
rightline=false,
leftline=false,
topline=false,
bottomline=false,
backgroundcolor=gray!10,
linecolor=gray,
innerleftmargin=5pt,
innerrightmargin=5pt,
innertopmargin=5pt,
innerbottommargin=5pt,
leftmargin=0cm,
rightmargin=0cm,
linewidth=4pt]{eBox}

\numberwithin{equation}{section}

\def\beq{\begin{equation}}
\def\eeq{\end{equation}}

\def\bea{\begin{eqnarray}}
\def\eea{\end{eqnarray}}

\def\beq{\begin{equation}}
\def\eeq{\end{equation}}
\def\bea{\begin{eqnarray}}
\def\eea{\end{eqnarray}}

\def\O{{\cal O}}

\def\Mpl{M_{\rm Pl}}
\def\Mp{M_{\rm Pl}}

\def\fnl{f_{\rm NL}}

\def\b{{\rm b}}

\def\k{{\vec k}}

\def\v{{\bf v}}
\def\x{{\vec x}}
\def\b{{\vec b}}
\def\y{{\vec y}}

\DeclareRobustCommand{\SkipTocEntry}[4]{}

\definecolor{blue3}{RGB}{31, 119, 180}
\definecolor{red3}{RGB}{	214, 39, 40}
\definecolor{orange3}{RGB}{255, 127, 14}
\definecolor{green3}{RGB}{44, 160, 44}

\newcommand{\then}{\quad \Rightarrow\quad}
\renewcommand{\O}{\mathcal{O}}

\newcommand{\ex}[1]{\langle #1 \rangle}
\newcommand{\R}{\mathcal{R}}
\renewcommand{\v}[1]{\vec{#1}}
\newcommand{\e}{\epsilon}

\begin{document}

\begin{titlepage}
\setcounter{page}{1} \baselineskip=16pt 
\thispagestyle{empty}



\begin{center}
{\fontsize{18}{18} \bf On the Symmetries of Cosmological Perturbations}
\end{center}

\vskip 20pt
\begin{center}
\noindent
{\fontsize{12}{18}\selectfont Daniel Green,$^1$ and Enrico Pajer$^2$}
\end{center}

\begin{center}

\vskip 8pt
\textit{$^1$  Department of Physics, University of California, San Diego, La Jolla, CA 92093, USA}

\vskip 8pt
\textit{$^2$  Department of Applied Mathematics and Theoretical Physics, Centre for Mathematical Sciences,
University of Cambridge, Wilberforce Road, Cambridge CB3 0WA, UK}
\end{center}

\vspace{0.4cm}
 \begin{center}{\bf Abstract}
 \end{center}
 
 \noindent

The space of inflationary models is vast, containing wide varieties of mechanisms, symmetries, and spectra of particles.  Consequently, the space of observational signatures is similarly complex. Hence, it is natural to look for boundaries of the space of models and their signatures.  In this paper, we explore the possible symmetries associated with the primordial cosmological perturbations and their correlators in the asymptotic future. Assuming the observed homogeneity, isotropy and (approximate) scale invariance, we prove three main results.  First, correlation functions of scalar metric fluctuations are uniquely characterized by soft theorems and are free from ambiguity under field redefinitions. Second, whatever the particle content and interactions, when the standard soft theorems apply, invariance under de Sitter boosts (linearly realized conformal invariance) is only possible if all connected correlators vanish identically, i.e.~if the theory is free.  Third, conformal invariance is the largest set of linearly realized (bosonic) symmetries of the correlators of any single scalar, irrespectively of any soft theorems or particle content.

\end{titlepage}

\restoregeometry

\newpage
\setcounter{tocdepth}{2}
\tableofcontents

\newpage

\section{Introduction}

Determining the nature of the universe at its earliest moments is one of the central goals of modern cosmology.  Observations strongly suggest inflation, or some other mechanism, was needed to produce superHubble fluctuations~\cite{Hu:1996yt,Spergel:1997vq,Dodelson:2003ip}.  Yet, the physics of the very early universe is hidden from us by the hot big bang, making it difficult to reconstruct the history of this epoch directly. The pattern of primordial density fluctuations that seed the growth of structure does give us hints about the earliest moments  but there is a seemingly endless list of possible mechanisms that can be made consistent with current observations.

While primordial perturbations do not directly reveal the time evolution, the symmetries relevant to the early universe do manifest themselves in their correlators.  Noether's theorem famously shows that symmetries constrain the allowed time evolution, forcing the local conservation of charge.  Equally profoundly, Coleman and Mandula~\cite{Coleman:1967ad} showed that in Minkowski space, the possible symmetries are constrained by self-consistent dynamics capable of producing a non-trivial S-matrix.  It is natural to wonder to what degree the possible symmetries of the early universe might be limited by the self-consistency of cosmological correlators. 

Cosmological correlators need not obey the Coleman-Mandula theorem (or generalizations thereof) for two reasons.  First, both Lorentz boosts and time-translations are spontaneously broken in cosmological models due to the background geometry and the vacuum expectation values of the fields that drive the time evolution. This breaks the important assumption of Lorentz invariance that is omnipresent in the S-matrix program. Moreover the scale of this breaking can be (and often is) parametrically larger than Hubble and so the effective theory can become strongly coupled \textit{before} Lorentz invariance is recovered in the flat space (short distance) limit. Second, unlike scattering ampitudes, correlation functions are not invariant in general either under field redefinitions or under total time derivatives. This implies that additional information must be provided about correlators to define them uniquely. As we will see, cosmology provides us with solutions to both of these problems and we will be able to find a cosmological analogue of the Coleman-Mandula theorem for the scalar correlators relevant to most models of the early universe.

Current cosmological observations indicate the primordial curvature fluctuations are approximately scale invariant.  It is natural to consider the limit where scale invariance becomes an exact symmetry and ask what set of additional symmetries could be realized in the same limit. To gain intuition, it is therefore useful to understand the origin of scale invariance in inflationary models.  At a most basic level, scale invariance arises from a time-translation symmetry, which ensures that each mode experiences the same history. While time-translations are a symmetry of flat space, they are spontaneously broken during inflation due to the time-evolution of the background, perhaps through the evolution of a scalar field $\phi(t)$.  However, it is possible to require the existence of a new (approximate) time translation symmetry. Such a symmetry can emerge as a diagonal combination of the original spontaneously broken time-translation symmetry and an additional internal global symmetry, which characteristically takes the form of a shift symmetry\footnote{It is important to stress that the presence of a shift symmetry does not per se imply the existence of this unbroken diagonal time-translation symmetry, as discussed in detail in \cite{Finelli:2018upr}. A simple counterexample is the Lagrangian $  P(X)-\lambda \phi $ in Minkowski. Rather, the shift symmetry implies an infinite series of recursive relations for the time dependence of the EFT parameters \cite{Finelli:2018upr}, as well as a new set of soft theorems \cite{Finelli:2017fml}. Hence, to obtain a scale invariant set of correlators a diagonal time-translation symmetry needs to be imposed \textit{in addition to} the existence of a shift symmetry.}  $\phi \to \phi+c$, ~\cite{Baumann:2011su,Baumann:2019ghk}. Intuitively, a shift symmetry ensures that perturbations are only sensitive to the background value of $  \dot\phi $, as higher derivatives can be eliminated using the background equations of motion. If we further assume that $  \dot \phi $ is approximately constant in time, perturbations must then be symmetric under time-translations. Formally, in the flat space limit, the conserved current $  t^{\mu} $ associated with the diagonal time-translations takes the form
\beq\label{eq:diagonal}
t^{\mu}=T^{0 \mu}+j^{\mu} \ ,
\eeq
where $T^{\mu \nu}$ is the stress tensor and $j^\mu$ is the (approximately) conserved current associated with the (approximate) shift symmetry.  

The most natural extension of a scale transformation would be to have a linearly-realized conformal symmetry.  Conformal invariance is famously realized as the result of the isometries of de Sitter spacetime but has its origins as the boosts in flat space, which are implemented by the current
\beq
K^{\mu 0 \lambda}=x^{\mu} T^{0 \lambda}-x^{0} T^{\mu \lambda} \ .
\eeq
Due to the time evolution of the background, these symmetries are necessarily broken (spontaneously) along with the time translations generated by $T^{00}$.  However, if we imagine there is an unbroken diagonal boost symmetry analogous to  \eqref{eq:diagonal}, we would required a higher spin current $j^{\mu \lambda}$ to combine with $K^{\mu \nu \lambda}$ to produce an unbroken symmetry.  Such a possibility is difficult to arrange in an interacting theory, as a linearly realized higher spin-symmetry would be forbidden by the Coleman-Mandula theorem.  Internal Galilean transformations~\cite{Nicolis:2008in,Burrage:2010cu,Creminelli:2010qf}, or other non-linearly realized symmetries, would still be permitted by this argument but the known examples have been shown to break the de Sitter isometries by explicit calculation~\cite{Deffayet:2009wt,Creminelli:2012ed}.  These considerations suggest that conformal invariance is severely restricted in inflationary models, a fact that we make precise in the rest of this work.

Symmetries can act linearly on small perturbations, hence being linearly realised, or they can be non-linearly realised as it occurs in spontaneous symmetry breaking. Linearly realized symmetries are easier to constrain because they separately restrict the possible functional form that each correlator can take. On the other hand, in cosmology we know we have a non-trivial gravitational background, which accounts for the expansion of the universe, and some additional non-trivial background of matter fields. Both generically spontaneously break time translations and boosts but can also break other symmetries. The ensuing non-linearly realized symmetries imply Ward-Takahashi identities that relate correlators with a different number of fields. This is harder to study because all correlators need to be solved for at once. Two approaches have been put forward in the literature to explore the landscape of non-linearly realized symmetries in cosmology. The first is the study of soft theorems, which employs the developments in \cite{Maldacena:2002vr,Creminelli:2004yq,Hinterbichler:2012nm,Assassi:2012zq,Hinterbichler:2013dpa,Creminelli:2012ed,Pajer:2017hmb,Hui:2018cag} to study residual diffeomorphisms. Additional non-linearly realized symmetries lead to a generalization of Weinberg's adiabatic modes \cite{Weinberg:2003sw} and new associated soft theorems, as for example for a shift symmetry \cite{Finelli:2017fml,Finelli:2018upr} or solid inflation \cite{Endlich:2012pz,Endlich:2013jia,Bordin:2017ozj,Pajer:2019jhb}. The second approach is a brute force classification of all possible symmetric Lagrangians. For a single scalar this was achieved in \cite{Pajer:2018egx,Grall:2019qof}, where a complete classification of all symmetries, irrespectively of their realization, was derived. In this work we will exclusively discuss linearly realized symmetries and comment on possible extensions in the conclusions.

 
\subsection{Summary of the results}\label{ssec:}

In this paper, we prove three theorems about primordial correlators under the assumption that they are all homogeneous, isotropic and scale-invariant:
\begin{itemize}[label=]
\item \textbf{Theorem 1:} In an attractor single-clock cosmology, without any further assumption on the particle content, the symmetries associated with adiabatic modes and the ensuing soft theorems uniquely fix the definition of curvature perturbations. In simpler terms, a set of scalar correlators are correlators of $  \zeta $ if and only if they satisfy all the soft theorems that generalize Maldacena's consistency relation \cite{Maldacena:2011jn}. This result proves invaluable when studying cosmological correlators exclusively based on their asymptotic values, i.e.~on the boundary, rather than following their time evolution, i.e.~in the bulk, as we do in this paper and in analogy with the S-matrix program. 
\item \textbf{Theorem 2:} In an attractor single-clock cosmology, without any further assumption on the particle content, linearly-realized conformal invariance (i.e.~invariance under de Sitter boosts) of $  \zeta $ may only arise in a free theory. As a consequence, any approach that crucially employs de Sitter isometries, such as those recently proposed in~\cite{Mata:2012bx,Arkani-Hamed:2018kmz,Baumann:2019oyu,Sleight:2019mgd,Sleight:2019hfp}, in single-field inflation can only describe slow-roll suppressed correlators of $  \zeta $.\label{one}
\item \textbf{Theorem 3:}  The only additional (linearly-realized) symmetries that the correlators of a scalar field $ \phi  $ can display, without vanishing, are the special conformal transformations. This a very strong restriction also on models in which curvature perturbations are not of the single-clock type, but rather are related to some other ``isocurvature'' fields. \label{three}
\end{itemize}
The combination of the above results tells us something remarkable. Cosmological observations indicate that $  \zeta $ has a homogeneous, isotropic and approximately scale-invariant power spectrum and therefore satisfies the assumptions of Theorem 3. Then, the only additional linearly-realized symmetries that $  \zeta $ can display are special conformal transformations. But from Theorem 2 we know that if these additional symmetries were present, then $  \zeta $ could only have vanishing connected correlators in the slow-roll decoupling limit. Hence, it follows from our results that there are only two mutually exclusive possibilities for our universe: 
\begin{itemize}
\item Primordial perturbations enjoy more symmetries than those we have already observed. Then these additional symmetries must be special conformal transformations, and all primordial non-Gaussianities: (a) are slow-roll suppressed and hence very small and/or (b) violate the consistency conditions.
\item Primordial non-Gaussianities are not slow-roll suppressed and so can be large, and they satisfy the consistency conditions. Then primordial perturbations display the largest possible set of symmetries, namely homogeneity, isotropy and (approximate) scale invariance.
\end{itemize} 
This conclusion somewhat parallels what we know about amplitudes: the observed Poincar\'e invariance is the largest set of (linearly-realized, bosonic) spacetime symmetries that a theory with a non-trivial S-matrix can enjoy. 

Note that our theorems still permit any internal (global) symmetries and/or non-linearly realized symmetries, just as for the Coleman-Mandula theorem. Also, our results are weaker than the corresponding constraints on Conformal Field Theories (CFTs) in~\cite{Maldacena:2011jn} as our arguments still allow any symmetries that act as the identity on $  \zeta $, but not on other fields. Given that many of the techniques and terminology we use are borrowed from CFTs, it's important to emphasize that our results do not follow trivially from holography and, as far as we can tell, they do not have an obvious analogue in a putative CFT dual to de Sitter, as discussed further in Section \ref{holog}.  Naively, one might interpret Theorem 2 as the statement that the trace of the stress tensor vanishes in a CFT.  Instead, one can check that this not the case. Furthermore, even the above description of scale invariance has a peculiar holographic origin~\cite{Baumann:2019ghk}.

This paper is organized as follows: In Section~\ref{sec:symmetries}, we discuss the symmetry algebra relevant to curvature fluctuations and prove Theorem 1.  In Section~\ref{sec:free}, we present two proofs of Theorem 2, namely that linearly-realized conformal invariance and the single-field consistency conditions combined imply that all the connected cosmological correlation functions of $  \zeta $ vanish.  In Section~\ref{sec:linear}, we prove Theorem 3 by classifying all possible linearly-realized symmetries for a general scalar field, which might or might not satisfy some soft theorems. We conclude in Section~\ref{sec:conc}. 


\section{The symmetries of curvature perturbations in single-clock inflation}\label{sec:symmetries}

In this section, we will define the symmetries we will use to constrain the form of cosmological correlators. Consider the FLRW metric
\begin{align}
ds^2 &=-dt^{2}+a(t)^{2}dx^{i}\delta_{ij}dx^{j}\\
&=a(\tau)^{2}\left[ -d\tau^{2}+dx^{i}\delta_{ij}dx^{j} \right]\,.
\end{align}
We will be interested in (attractor) single-clock accelerated cosmologies, $ \ddot a >0 $, for which curvature perturbations eventually become longer than the Hubble radius and freeze out. In the asymptotic future we can neglect the time dependence and focus on the spatial dependence of equal-time correlators. We will phrase our discussion in terms of curvature perturbations on constant density hypersurfaces $  \zeta(\v{x}) $ at future infinity, which is what we can measure in cosmological observations
\begin{align}
\zeta(\v{x},\tau)\to \zeta(\v{x})\quad \text{as}\quad \tau \to  0\,.
\end{align} 
It should be noted that all of our results are also valid for curvature perturbations on comoving hyperfurfaces $  \R(\v{x}) $, since the difference between $\R$ and $\zeta$ vanishes at future infinity under our assumptions.

Around any FLRW cosmology, $\zeta$ obeys an SO$ (4,1)  $ symmetry associated with the isometry of spacetime and the residual large gauge transformations~\cite{Creminelli:2012ed,Hinterbichler:2012nm}. The ISO(3) subgroup represents the standard rotation and translation symmetries, which are linearly realized on $  \zeta $, with generators
\begin{align}\label{P}
P_{i}&: \delta \zeta=-\partial_{i}\zeta\,, \\
M_{ij}&: \delta \zeta=2x_{[i}\partial_{j]}\zeta\,.
\end{align}
At future infinity, when the subleading time-dependence of $  \zeta $ can be neglected, there are also four more symmetries that act non-linearly on $  \zeta $ . These act like three-dimensional euclidean conformal transformations and are generated by non-linear dilations and special conformal transformations, with generators given by
\begin{align} 
D_{\rm NL}&: \delta \zeta  =-1-\x \cdot \vec \partial_\x \zeta\,, \\ 
K_{\rm NL}^{i}&: \delta \zeta =-2 x^{i}-2 x^{i}\left(\x \cdot \vec \partial_{\x} \zeta\right)+x^{2} \partial^{i} \zeta\,.
\end{align}
The first term in each of these two generators is a (non-linear) shift, while the remaining terms in $  K_{\text{NL}} $ are the usual conformal transformations of a scalar field of zero conformal dimension. All of the above symmetries are present for any single-clock attractor model of inflation.

In addition to these mandatory symmetries, models of inflation that are phenomenological viable also display an approximate linearly-realized dilation symmetry, which is responsible for the observed approximate scale invariance of primordial perturbations. This symmetry can be thought of as arising from a rescaling of the coordinates
\begin{align}
x^i \to (1+\lambda) x^i  \,,
\end{align}
under which operators transform according to the their conformal dimension $  \Delta_{\O} $. For scalar operators $  \O $ one has
\begin{align}
{\cal O}(\x) \to (1+\lambda)^{\Delta_{\O}} {\cal O}((1+\lambda) \x)\,.
\end{align}
We can define the associated infinitesimal generator $  D $ by
\begin{align}
D&:  \delta \O(\x) = -   \Delta_{\O}{\cal O}(\x)  - \x \cdot  \vec \partial {\cal O}(\x)  \,.
\end{align}
The conformal dimension $ \Delta_{\zeta} $ of $  \zeta $ actually has to vanish in single-field attractor inflation. To see this, we notice that 
\begin{align}
[D,D_{\text{NL}}]=\Delta_{\zeta}\,,
\end{align}
and therefore, if $ \Delta_{\zeta} \neq 0$, we can define a new generator $ S$ that acts on $ \zeta$ as a shift symmetry
\begin{align}
S\equiv  \frac{1}{\Delta_{\zeta}}[D,D_{\text{NL}}]:\quad \delta \zeta=1\,.
\end{align}
Consequently, $ D_{\text{NL}}+S$ must also be a symmetry and acts linearly as 
\begin{align}
D_{\text{NL}}+S: \quad \delta \zeta = -x^{i}\partial_{i}\zeta\,.
\end{align}
This is nothing but a dilation with $ \Delta_{\zeta} =0$. We found that, if $  \Delta_{\zeta}\neq 0 $, the theory would have to be simultaneously invariant under two types of linearly-realised dilations in which $ \zeta $ has two different conformal dimensions, namely zero and $  \Delta_{\zeta} $. No correlator can be invariant under this set of symmetries\footnote{When scale invariance is broken, the quadratic action may still be usefully described in terms of a scaling symmetry with $\Delta_\zeta \neq 0$~\cite{alberto}.  We have showed that such a symmetry is never exact.  } and such a theory cannot exist\footnote{For a less formal but more explicit argument, recall that dilation invariance reduces to the constraint
\begin{align}
-3+\sum_{a}(3-\Delta_{\zeta})+\v{k}_{a}\cdot \partial_{\v{k}_{a}} B_{n}=0\then  P\propto k^{-3+2\Delta_{\zeta}}\,, \quad B_{3}\propto k^{-6+3\Delta_{\zeta}} \,.
\end{align}
But then Maldacena consistency relation gives $  B_{3}\propto P^{2} $, which cannot be satisfied unless $  \Delta_{\zeta}=0 $.}. Notice that for this argument we did not have to invoke $ K^{\text{NL}}_{i}$. From now on we will assume $ \Delta_{\zeta} =0$. 

We will be interested in answering the following question: Can there be additional symmetries beyond the ones we discussed above? Our Theorem 2 and 3 say that, under different assumptions, the answer is negative. For example, one might try to add linearly-realized de Sitter boosts as a symmetry of $  \zeta $. This is precisely the symmetry that a spectator field in de Sitter would enjoy. When acting at future infinity, such a transformation takes the form of a three-dimensional euclidean special conformation transformation,
\beq
x^i \to x'^{i}= x^i + b^i x^2 - 2 x^i \vec b \cdot \x \ .
\eeq
Operators of conformal dimension $\Delta$ then transform as 
\beq
\qquad {\cal O}(\x) \to (1-2 \b\cdot \x)^\Delta {\cal O}(\x')\,,
\eeq
but we will only need the linearized form of these transformations,
\begin{align}
K^i&:  \delta \O(\x) =  \left[ - 2 \Delta x^i + x^2 \partial^i - 2 x^i \x \cdot  \vec \partial \right] {\cal O}(\x) \,.\label{eq:SC}
\end{align}

 
\subsection{The symmetry algebra}\label{ssec:algebra}

It will be useful to take a closer look at the set of generators we have just defined, in real as well as in Fourier space:
\begin{align} \label{syms}
P_{i}&: \delta \zeta=-\partial_{x^{i}}\zeta\,, & \delta \zeta&= -ik_{i}\zeta\,, \\
M_{ij}&: \delta \zeta=2x_{[i}\partial_{x^{j]}}\zeta\,, & \delta \zeta &= 2 k^{[i}\partial_{k^{j]}}\zeta\,,\\
D&: \delta\zeta  =-\x \cdot \vec \partial \zeta\,, & \delta \zeta &=\left[ 3+k^{i}\partial_{k^{i}} \right]\zeta\,, \label{3b} \\ 
K^i&:\delta \zeta  =-2 x^{i}\left(\x \cdot \vec \partial \zeta\right)+x^{2} \partial_{x^{i}} \zeta\,, & \delta \zeta&= i\left[ 2 \vec{k} \cdot \vec{\partial} \partial_{k^{i}}-k_{i}\partial^{2}+6 \partial_{k^{i}} \right]\zeta\,, \\
 D_{\rm NL}&: \delta\zeta  =-1-\x \cdot \vec \partial  \zeta \,,\\ 
K_{\rm NL}^{i}&:\delta \zeta =-2 x^{i}-2 x^{i}\left(\x \cdot \vec \partial  \zeta\right)+x^{2} \partial^{x^{i}} \zeta  \,.\label{syms2}
\end{align}
It turns out that these generators by themselves do not close to form an algebra, a fact that will play a key role in our second proof of Theorem 2 in Section \ref{ope}. To see this, it is more convenient to use the following linear combinations of generators
\begin{align}\label{Q}
Q\equiv D-D_{\text{NL}}&: \delta \zeta=1  \,, \\ 
V_{i}\equiv \frac{1}{2} \left( K_{i}-K_{i}^{\text{NL}} \right)&: \delta \zeta =x_{i}\,,
\end{align}
which we recognize as the generators of a (euclidean) Galilean symmetry \cite{Nicolis:2008in}. By direct computation we find
\begin{align}
[V_{l},K_{i}]&=\left( -2x_{i}x_{l}+x^{2}\delta_{il} \right)\,.
\end{align}
Since the commutator of two symmetry generators must also be a symmetry generator, we need to add to the algebra of symmetries $  V_{li}  $ defined by
\begin{align}\label{2}
V_{li}&: \delta \zeta=\left( -2x_{i}x_{l}+x^{2}\delta_{il} \right) \,.
\end{align}
But we cannot stop here. Commuting $  V_{ij} $ once more with $  K_{l} $ we find that we need to include yet another set of generators, 
\begin{align}\label{3}
V_{ijl}\equiv[V_{ij},K_{l}]&: \delta \zeta= 8x_{i}x_{j}x_{l} -2x^{2} \left(x_{l}\delta_{ij}+x_{i}\delta_{jl}+x_{j}\delta_{il} \right)\,. 
\end{align}
This continues ad infinitum. As we commute $  V_{ijl} $ with $  K_{m} $ more and more times, we are forced to include generators with increasing powers of $  x $. We could have seen the need for the infinitely many additional generators from a more formal argument. From Jacobi identities, or simply by direct calculation we know that
\begin{align}\label{31}
[[A,B],D]=\left(  \Delta_{A}+\Delta_{B}\right)[A,B]\,,
\end{align}
for any two generators $  A $ and $  B $ with conformal dimension $  \Delta_{A,B} $ defined as
\begin{align}
[A,D]=\Delta_{A}A\,,
\end{align}
and similarly for $  B $. But $  \Delta_{V_{i}}=\Delta_{K_{i}}=1 $ and so their commutator must have $  \Delta_{V_{ij}}=2 $, which is indeed what we see in \eqref{2}. Every time we commute with $  K_{i} $ we find a new generator with conformal dimension increased by one. These additional symmetries induce transformations on $  \zeta $ of the schematic form (indices are implicit)
\begin{align}
V_{(n)}\equiv V_{i_{1}\dots i_{n}}: \delta \zeta \sim x^{n}\,.
\end{align}
It will be useful to notice that the $  V_{(n)} $ are totally symmetric in all their indices. For $  V_{ij} $ and $  V_{ijl} $ this can be seen from their explicit expressions in \eqref{2} and \eqref{3b}. For all the higher order generators we can provide the following proof by induction. Assume $  V_{(n)} $ is totally symmetric in its $  n $ indices. Then
\begin{align}
V_{(n+1)}=V_{i_{1}\dots i_{n-1}i_{n}}&=[V_{(n)},K_{i_{n}}]=[[V_{n-1},K_{i_{n-1}}],K_{i_{n}}]\\
&=[V_{n-1}K_{i_{n-1}},K_{i_{n}}]-[K_{i_{n-1}}V_{n-1},K_{i_{n}}]\\
&=[V_{n-1},K_{i_{n}}]K_{i_{n-1}}-K_{i_{n-1}}[V_{n-1},K_{i_{n}}]\\
&=[[V_{n-1},K_{i_{n}}],K_{i_{n-1}}]=V_{i_{1},\dots i_{n}i_{n-1}}\,,
\end{align}
where we used the fact that $  [K_{i},K_{j}]=0 $. The above results shows that $  V_{n+1} $ is symmetric in its last two indices. But from its definition, $  V_{(n+1)} $ is also symmetric in its first $  n $ indices and so it must be totally symmetric. In particular, the number of components of $  V_{(n)} $ is
\begin{align}
\begin{pmatrix} 3+n-1  \\ n \end{pmatrix}=\frac{(2+n)!}{n!\, 2!}=\frac{1}{2}(2+n)(1+n)\,.
\end{align}


\subsection{Theorem 1: field redefinitions}

In this paper, we are deriving constraints on cosmological correlators based exclusively on their symmetries, without any reference to an underlying Lagrangian. One immediate concern with this approach is that correlation functions are not invariant under field redefinitions\footnote{Without loss of generality, when discussing correlators at future infinity we can disregard the many field redefinitions that vanish in that limit. This is in contrast to approaches that follow the time evolution in the ``bulk'', where all field redefinitions are used to simplify the Lagrangian, as e.g.~in \cite{Bordin:2017hal,Bordin:2020eui}.}. For example, non-vanishing connected correlators may result simply from a field redefinition of a free theory in which all connected correlators vanish. This is in contrast to what happens in flat space, where we have a natural and unambiguous definition of a free theory: the S-matrix must be the identity. This flat-space definition is invariant under (perturbative) field redefinitions, as can be seen from the LSZ reduction formula. 

The issue of field redefinition is common to any attempt to characterize cosmological correlators from a ``boundary'' perspective, namely from their value in the asymptotic future where they approach some constant, without explicit reference to a ``bulk'' theory, i.e.~to the time evolution when the modes in the correlator are sub-Hubble. Specifically, if one finds some way to compute the correlators of some scalar, without reference to a Lagrangian, how does the correlator know which scalar it is supposed to compute? For a trivial example, one can add to a field $  \phi $ a ``local'' term, $ \phi \to \phi+ \phi^{n} $, which changes the correlator. More interestingly, one can also add arbitrary space and time derivatives, with appropriate powers of the scale factor and/or inverse Laplacians so that the field redefinition does not vanish at spatial infinity. Again the correlator changes. 

In this section, we point out that if one is interested in the correlators of $  \zeta $ in single-clock inflation, as it is often the case in cosmology, there is a simple way to address the ambiguity induced by field redefinitions. In particular, we will prove that in single-field inflation \textit{a given correlator is a correlator of $  \zeta $ if and only if it satisfies all the soft theorems \cite{Maldacena:2002vr,Creminelli:2004yq,Hinterbichler:2012nm,Assassi:2012zq,Hinterbichler:2013dpa,Creminelli:2012ed,Pajer:2017hmb} that enforce on the correlators the symmetries $  D_{\text{NL}} $ and $  K_{\text{NL}} $} . We will refer to this result as Theorem 1. There are two proofs of this theorem. The first proof will be presented in this section. The second proof follows exactly the same steps as in the discussion of the Operator Product Expansion (OPE) of Section \ref{ope}, and is left to the interested reader as an exercise. 

Broadly speaking, our goal is to define operators by how they transform under the mandatory symmetries discussed in this section.  Of course, the symmetries alone do not uniquely define the operators of a theory, as the ambiguity under field redefinitions cannot vanish entirely.  Instead, the field redefinitions need not commute with the symmetry, in the sense that the symmetries will act differently on operators in theories related by a field redefinition.  Often, there is a particularly convenience choice for representation of the operators that simplifies the action of the symmetry on the operators. The canonical definition of $\zeta$ is such an example\footnote{When defined in terms of the metric, $ds^2 = -dt^2 + a^2(t) e^{2\zeta}d\x^2$, we see that a constant shift of $\zeta$ is related to rescaling $\x$.  }, as this choice simplifies the non-linearly realized scale transformations. If this basis of operators is unique, then we can define the ``boundary" correlation functions by the transformation of the operators under these symmetries. We will now demonstrate that $\zeta$ is uniquely defined in this way.

For field redefinitions of $\zeta$, we will only need the following generators:
\begin{align}
D_{\rm NL}&: \delta \zeta  =-1-\x \cdot \vec  \partial  \zeta \,,\\
K_{\rm NL}^{i}&:\delta \zeta =-2 x^{i}-2 x^{i}\left(\x \cdot \vec \partial \zeta\right)+x^{2} \partial^{i} \zeta \,, \\
D&: \delta\zeta  =-\x \cdot \vec \partial \zeta\,.
\end{align}
We will show that this set of generators uniquely determines $\zeta$.  Furthermore, the first two generators are present in any inflationary model and the third is typically an approximate symmetry (as suggested by observations).  

Let's start with field redefinitions of $  \zeta $ involving any operator that is the sum or product of $  \zeta $ at the same point, without any derivatives\footnote{In the cosmological literature, this type of transformation goes sometimes under the name of ``local'' redefinitions. We avoid using this terminology here because in quantum field theory a ``local operator'' usually refers to sum and product of operators \textit{as well as} their derivatives at the same point. We will use the word ``local'' with this second meaning in mind, in line with the QFT literature.},
\beq\label{F}
\tilde \zeta(\x) =\zeta(\x) + F(\zeta(\x)) \,,
\eeq  
where $  F $ is a real function, $  F:\mathbb{R}\to\mathbb{R} $. For $\tilde \zeta$ to transform in the same way as $\zeta$ under $  D_{\rm NL}$, we need to require
\beq
\frac{\partial F(\zeta(\x) )}{\partial \zeta(\x)}= 0\,.
\eeq
Since this must hold for any value of $  \zeta(\x) $, we conclude that $F(\zeta)$ must be a constant function. A constant shift $ \zeta \to \zeta + C $ doesn't change any connected correlator and it is usually fixed by demanding that perturbations have zero expectation value, which implies $  C=0 $. Thus, even without using $  D $, we see that field redefinitions of the form \eqref{F} are fixed by $D_{\rm NL}$. This is also straightforward to see at the level of correlators, where the field redefinition in \eqref{F} generates terms that violate the consistency conditions to leading order.

Now suppose we included derivatives of $\zeta$ as part of our redefinition, e.g.
\beq
\tilde \zeta(\x) =\zeta(\x) + F(\nabla^2 \zeta(\x)) \ .
\eeq  
Because $\nabla^2 \zeta$ transforms linearly under $D_{\rm NL}$, $\tilde \zeta$ and $  \zeta $ transform the same under the non-linear part of $  D_{\rm NL} $, namely $  D_{\rm NL}-D $.  However, now we look at the symmetry generated by $D$. The scaling dimension of $\nabla^2 \zeta$ clearly satisfies $\Delta_{\nabla^2 \zeta}=2  >0$.  In contrast, $\zeta$ has dimension $\Delta_\zeta = 0$. If we then impose that $\tilde \zeta $ has the same dimension of $  \zeta $ under $  D $, namely $  \Delta_{\zeta}=\Delta_{\tilde \zeta}=0 $, then we need to impose
\beq
F(\nabla^2 \zeta) = F(\lambda \Delta_{\nabla^2 \zeta} \nabla^2 \zeta)\,,
\eeq
for some real parameter $  \lambda $. Since this must be true for any $  \nabla^2 \zeta $ and small but non-vanishing $  \lambda $, we conclude that $F(\nabla^2 \zeta) $ must be a constant function. The same will be true of any function of non-zero dimension. In single-clock inflation, there are no local operators of dimension zero other than $\zeta^n(\x)$, which we already argued is not allowed in the field redefinition. We conclude that $  D_{\text{NL}} $ and $  D $ fix the redefinition of $\zeta$ by any local operators. 

Thus far, we have not addressed the possibility of a field definition by a non-local operator. To begin, let's write a general non-local field redefinition as
\beq\label{nloc}
\tilde \zeta(\x) = \zeta(\x) + \left(\prod_{i =1}^N \int d^3 x_i {\cal O}(\x_i) \right) G(\{\x-\x_i \}, \{ \x_i-\x_j \} )\,,
\eeq
where ${\cal O}(x_i)$ are some operators build from the sum and product of $  \zeta $ and its derivatives at the same point and $  G $ is some kernel. From our previous discussion, we already know that each $  \zeta $ needs to appear with at least one derivative, in order to have a chance to not spoil the transformation generated by $  D_{\text{NL}} $. For example, consider the field redefinition
\beq\label{eq:nonlocal_eg}
\tilde \zeta = \zeta+ \frac{1}{\nabla^2}\left( \partial_i \zeta \partial^i \zeta \right) \ .
\eeq
This non-local redefinition avoids the constraints we derived previously from $  D_{\text{NL}} $ and $  D $ because of two reasons: (1) the derivatives acting on $\zeta$ ensure the action of $  D_{\text{NL}} $ is the same as that of $  D $ and (2) the inverse Laplacian allows the scaling dimension to remain zero, hence leaving the transformation induced by $  D $ unchanged. Of course the price we pay is that the inverse Laplacian makes the operator non-local. Such form of non-locality are not a priori problematic. Indeed inverse Laplacians emerge generally when solving the ADM constraints for the $  g_{0\mu} $ components of the metric, and the calculation of the bispectrum in single field inflation in \cite{Maldacena:2002vr} employs a similarly non-local field redefinition. 

Under dilations, \eqref{nloc} transforms as
\bea
D(\tilde \zeta(\x) ) &=& \zeta(\lambda \x) +  \left(\prod_{i =1}^N \int d^3 x_i \lambda^{\Delta_i} {\cal O}(\lambda \x_i) \right) G(\{\x-\x_i\}, \{ \x_i-\x_j \} ) \\
&=&  \zeta(\lambda \x) +  \left(\prod_{i =1}^N \int d^3 y_i \lambda^{\Delta_i-3} {\cal O}(\y_i) \right) G(\{\x-\y_i/\lambda \}, \{ (\y_i -\y_j)/\lambda \} ) \ ,
\eea
where $  \Delta_{i} $ are the scaling dimensions of the operators $  \O_{i} $ and we used translation invariance to fix the form of our kernel, $G(\{\x-\x_i\}, \{ \x_i-\x_j \} )$.  For $ \tilde \zeta $ to transform in the same way as $  \zeta $, namely $D(\tilde \zeta(\x) ) = \tilde \zeta(\lambda x)$, we require
\beq
\lambda^{\sum_i (\Delta_i -3)} G(\{\x-\y_i/\lambda \}, \{ (\y_i -\y_j)/\lambda \} ) = G(\{\lambda \x-\y_i \}, \{ \y_i -\y_j \} )\,.
\eeq
This fixes the scaling behavior of our kernel but is easily satisfied. In our simple example in \eqref{eq:nonlocal_eg} $,{\cal O}(x_1) = \partial_i \zeta(x_1) \partial^i \zeta (x_1)$ and $G(\x-\x_1)= 1/|\x-\x_1|$ is the Green's function that implements the inverse Laplacian.  Since ${\cal O}(x_1)$ has $\Delta_1 =2$, we check that 
\beq
\frac{1}{\lambda  } G(\x-\y_1/\lambda) = G(\lambda \x - \y_1) \,,
\eeq
as required to match the scaling behavior.

Let's see what we can learn from imposing that the transformation generated by $K_{\rm NL}$ remains unchanged. First of all, the non-linear part of the transformation involves a shift of $  \zeta $ by $  x^{i} $. By an argument precisely analogous to that around \eqref{F}, this implies that the operators $  \O $ in the field redefinition need to have no less than two derivates acting on $ \zeta  $. If they do not, then it is straightforward to see that $\tilde \zeta$ does not transform the same way as $\zeta$ under $  K_{\text{NL}} $. So in the following we will assume that this is the case and therefore $  K_{\text{NL}} $ induces the same transformation as $  K $.

Now we turn to the constraints imposed by the linear action of $K_{\rm NL}$, namely
\bea\label{eq:Knl_field}
K_{\rm NL}(\tilde \zeta(\x) ) = K_{\rm NL} \zeta(\x) +  \left(\prod_{a =1}^N \int d^3 x_a \left| \frac{\partial \x_a'}{\partial \x_a} \right|^{\Delta_i/3} {\cal O}(\x'_a) \right) G(\{\x-\x_a\}, \{ \x_a-\x_{a'} \} )\,, 
\eea
where $\x' =\x + \vec b x^2 - 2 \x \vec b \cdot \x$. To proceed, it is convenient to change the variables of integration to $\y=\x'$ in \eqref{eq:Knl_field} to give
\bea\label{eq:Knl_y}
K_{\rm NL}(\tilde \zeta(\x) ) = K_{\rm NL} \zeta(\x) +   \left(\prod_{a =1}^N \int d^3 y_a \left| \frac{\partial \x_a}{\partial \y_a} \right|^{1-\Delta_a/3 } {\cal O}(\y_a) \right) G(\{\x-\x_a\}, \{ \x_a-\x_{a'} \} ) \ ,
\eea
where $  \x_{a}=\x_{a}(\y_{a}) $. In order for this to transform the same way as $\zeta(\x)$ we would need
\bea
K_{\rm NL}(\tilde \zeta(\x) ) = K_{\rm NL} \zeta(\x) +  \left(\prod_{a =1}^N \int d^3 y_a {\cal O}(y_a) \right) G(\{\x+\vec b x^2 -2 \x \vec b \cdot \x -\y_a\}, \{ \y_a-\y_{a'} \} ) 
\eea
to transform like a local operator, or equivalently
\beq
\prod_{a =1}^N \left| \frac{\partial \x_a(y_a)}{\partial \y_a} \right|^{1-\Delta_a/3 } G(\{\x-\x_a(y_a)\}, \{ \x_a(y_a)-\x_{a'}(y_{a'}) \} ) = G(\{\x+\vec b x^2 -2 \x \vec b \cdot \x -\y_a\}, \{ \y_a-\y_{a'} \} ) \,.
\eeq 
We can Taylor expand this expression in $\vec b$ on both sides.  Matching at each order in $\b$, we will relate functions of $\y_a$ on the left-hand side to functions of $\x$.  For generic $\x$ and $\y_a$ (i.e.~$\x \neq \y_a$), the only viable solution is that both sides of these equations are constants, so $\Delta_a = 3 $ and $G(\{x-x_a \}, \{ x_a-x_{a'} \} ) $ is constant.  These operators are independent of $\x$ and thus are not of interest for correlation functions of modes with finite momenta.  Alternatively, the kernel can vanish except at discrete points where $\x = \y_a$ (i.e.~$\delta$-functions).  These solutions are just local operators and thus are excluded by our previous arguments.  

Summarizing, we have proven that there are no field redefinitions that leave the action of $  D_{\text{NL}} $ and $  K_{\text{NL}} $ invariant, and so these symmetries uniquely fix the definition of $  \zeta $. At the level of correlators, these symmetries induce an infinite number of soft theorems \cite{Maldacena:2002vr,Creminelli:2004yq,Hinterbichler:2012nm,Assassi:2012zq,Hinterbichler:2013dpa,Creminelli:2012ed,Pajer:2017hmb}, which collectively identify a given set of correlators as the correlators of $  \zeta $.


\section{Theorem 2: de Sitter invariance implies a free theory}\label{sec:free}

In this section, we will prove that, in the decoupling limit, exact conformal invariance is only possible when $\zeta$ is a purely Gaussian field\footnote{A similar conclusion was drawn in~\cite{Creminelli:2012ed} using Galileons, but was limited to actions with two-derivative equations of motion and thus did not consider the generic higher derivative operators that appear in an EFT.}, a result to which we refer as Theorem 2.  We will first show this by a brute force application of the symmetries in their differential operator form.  We will then confirm and generalize these results using the Operator Product Expansion (OPE). 

Our main assumption in proving this result is that the correlators of $  \zeta $ are symmetric under the non-linearly realized symmetries generated by $  D_{\text{NL}} $ and $  K_{\text{NL}} $, and therefore satisfy the standard soft theorems that generalize Maldacena's consistency condition. As long as the soft theorems apply to $  \zeta $, both the particle content of the theory and the interactions can be arbitrary. For example, $  \zeta $ can have arbitrary interactions with massive particles of arbitrary spin. In particular, we don't assume that gravity is described by general relativity. Diff invariant theories of gravity that respect the standard soft theorems are also constrained by Theorem 2. 
Conversely, our conclusions do not apply when the soft theorems are invalid, such as for example in the presence of additional massless scalars whose fluctuations are eventually converted into adiabatic fluctuations. 

Before we proceed, let's observe that exact conformal invariance implies also exact scale invariance (by closure of the algebra for the $  [P_{i},K_{j}] $ commutator). Therefore, our interest in this section is in the behavior of $\zeta$ correlators in the \textit{scale-invariant decoupling limit}, namely
\begin{align}
\frac{H}{\Mpl},\,\frac{\dot H}{H^{2}},\,\frac{\dot H}{\Mpl^{2}} \to 0 \quad \text{with} \quad \frac{H^{4}}{\dot H\Mpl^2} \sim \text{const.} \quad \quad \text{(decoupling)}\,, \\
\e\equiv-\frac{\dot H}{H^{2}},\eta\equiv \frac{\dot \e}{\e H},\,\xi_{n\geq 3}\equiv \frac{\partial \log \xi_{n-1}}{H \partial t} \quad \to 0 \quad \quad \text{(scale invariant)}\,,
\end{align} 
in which gravitational interactions become negligible, while the amplitude of the primordial power spectrum remains fixed.
This limit is forced upon us because we found that scale invariance of the correlators is only possible when $\Delta_{\zeta} =0$.  This implies that the power spectrum of $\zeta$ takes the form $P(k)= A_s k^{-3}$ and hence $n_s-1=0$ along with any other deviations from exact scale invariance.  In the presence of dynamical gravity, $\epsilon = - \dot H /H^2 > 0$ and this will generically lead to small deviations from scale invariance.  However, in the decoupling limit, the geometry becomes pure de Sitter and exact scale invariance can be achieved, as long as all other Hubble slow-roll parameters $  \{\eta,\xi_{n}\} $ are also negligible.

An indirect consequence of this choice is that the metric degrees of freedom other than $\zeta$ will decouple from $\zeta$.  This includes both the tensor modes and the lapse and shift (i.e.~the $N$, $N^i$ ADM components).  We can see this most directly by working in a gauge where the scalar metric fluctuation is zero (flat gauge) and $\zeta$ is related to the fluctuations of a scalar field in a non-dynamical space-time.

\subsection{First proof: differential operators}\label{sec:diff}
It is most straightforward to see why non-Gaussian correlators vanish by acting with the symmetry generators on the connected correlation functions directly. We will demonstrate the idea of the proof working with the bispectrum, and generalize it later to higher point correlators\footnote{We have become aware of an unpublished manuscript by P.~Creminelli in which a similar argument based on differential operators was independently derived.}.  It is well-known that the position dependence of any three-point function is fixed by conformal invariance and the conformal dimensions and spins of the operators.  For a scalar operator of dimension $\Delta = 0$, like $\zeta$, the conformally invariant bispectrum takes the form \cite{Pajer:2016ieg}\footnote{If we insist on exact scale invariance then $  C_{\mathrm{con}}=0  $. But our argument applies more generally for any $  C_{\mathrm{con}}  $, so we will keep also this ``conformal shape'' in the following, since its variation under conformal transformations coincides with the local term.}
\beq\label{eq:conformal_bispectrum}
\langle \zeta_{\k_1} \zeta_{\k_2} \zeta_{\k_3} \rangle = \fnl^{\rm local}\frac{\sum_{a}k_{a}^{3}}{k_{1}^{3} k_{2}^{3} k_{3}^{3}}+ C_{\mathrm{con}} \frac{\log \left(K / k_{*}\right) \sum_{a=1}^{3} k_{a}^{3}-\sum_{a \neq b} k_{a}^{2} k_{b}+k_{1} k_{2} k_{3}}{k_{1}^{3} k_{2}^{3} k_{3}^{3}}\,,
\eeq
where $K=k_1+k_2+k_3$.  

We would now like to show this is consistent with the non-linearly realized conformal transformations only if $  \fnl^{\rm local}=C_{\mathrm{con}}=0 $.  One consequence of $   D_{\text{NL}}  $ and $  K_{\text{NL}} $ are single-field consistency condition, which fix the leading and next-to-leading order coefficients in the soft limit of all correlators in terms of lower order correlators. As it is well known, the consistency conditions can be invalidated by the presence of additional massless scalar fields or when the background evolution is not an attractor \cite{Namjoo:2012aa}, and so our proof does not apply to those cases. 

In the slow-roll decoupling limit where $n_s-1 = 0$, the consistency condition implies that when we take one momentum to be soft, say $ k_1 \to 0$, the bispectrum takes the form 
\beq\label{eq:sfcc}
\lim_{\k_1 \to 0} \langle \zeta_{\k_1} \zeta_{\k_2} \zeta_{\k_3} \rangle  =  P(k_1) P(k_2) (0 + 0 \times k_1+ {\cal O}(k_1^2)) \ .
\eeq
In other words, the leading scaling with $k_1 \to 0$ is at most $k_1^{-1}$.  Expanding the conformally invariant answer, the scaling in the soft limit is $k_1^{-3}$ unless both $\fnl^{\rm local} = 0$ and $C_{\mathrm{con}}= 0$.  The latter may not be obvious so we show it explicitly:
\beq
\lim_{\k_1 \to 0} \langle \zeta_{\k_1} \zeta_{\k_2} \zeta_{\k_3} \rangle = \frac{C_{\rm con}}{k_1^3 k_2^6} \left[  \log (2 k_2 / k_\star) (2 k_2^3) - 2 k_2^3  \right]\,,
\eeq
which violates the consistency conditions if $C_{\rm con} \neq 0$.  Furthermore, because of the log, we cannot cancel the two terms against each other to get a result that vanishes in the squeezed limit.  We see that by a brute force application of the set of symmetries $ \{ D , K,  D_{\text{NL}} ,  K_{\text{NL}} \}$, the only consistent bispectrum is a vanishing one.

We would like to extend the above argument to all connected $N$-point function of $\zeta$. Unlike the bispectrum, higher point functions are not uniquely determined by conformal invariance. Fortunately, it turns out that the tension between the linear and non-linearly realized conformal transformations is evident from the squeezed-limit alone.  Our strategy is therefore to eliminate the need to determine the full shape of a given $N$-point function by working directly with the squeezed limit.  Specifically, we will use generators of the symmetry in terms of differential operators in Fourier space,  
\bea
K^i {\cal O}(\vec k) &\to& \left[ - 2 i \Delta  \partial_{k_i} - i  \partial^2_{\vec k}  k^i +i 2  \partial_{k_i}  \partial_{\vec k} \cdot  \vec k \right] {\cal O}(\k) \\
&=&i \left[ - 2 \Delta \partial_{k_i} -  k^i  \partial^2_{\vec k} +6 \partial_{k_{i}}+2 ( \vec k \cdot  \partial_{\k }) \partial_{k_i }\right] {\cal O}(\k) \ ,
\eea
which are obtained directly from transforming \eqref{eq:SC}.  We will demand that the scaling behavior in the limit of one vanishing momentum is consistent with these symmetries.

Let's first repeat the argument for the bispectrum.  The bispectrum is only a function of $k_{a} = | \vec k_a |$ for $  a=1,2,3 $, and is therefore conformally invariant when 
\beq
\vec b \cdot \vec K \langle \zeta_{\k_1} \zeta_{\k_2} \zeta_{\k_3} \rangle = \sum_{a=1}^3 \vec b \cdot \k_a \left[ \frac{\partial^2}{\partial k_a^2} + \frac{4}{k_a} \frac{\partial}{\partial k_a}  \right] \langle \zeta_{\k_1} \zeta_{\k_2} \zeta_{\k_3} \rangle = 0 \ .
\eeq 
Now let's take the limit $k_1 \to 0$ and define $B(k_1,k_2,k_3)= \langle \zeta_{\k_1} \zeta_{\k_2} \zeta_{\k_3} \rangle'$, where a prime indicates that we dropped a factor of $  (2\pi)^{3}\delta_{D}^{3}(\sum_{a}\k_{a}) $.  Let's assume that the leading power of $k_1$ in the soft limit is $k_1^\alpha$ for some $  \alpha $, so that 
\beq
\lim_{k_1 \to 0} B(k_1,k_2,k_3) = k_1^\alpha F(k_2,k_3) + {\cal O}(k_1^{\alpha +1})\,.
\eeq
Applying the differential operator we have
\beq\label{eq:bi_constraint}
\vec b \cdot \vec k_1  \left[ \alpha (\alpha-1) + 4 \alpha \right] k_1^{\alpha-2} F(k_2,k_3) = - k_1^{\alpha}  \sum_{a=2}^3 \vec b \cdot \k_a \left[ \frac{\partial^2}{\partial k_a^2} + \frac{4}{k_a} \frac{\partial}{\partial k_a}  \right] F(k_2,k_3)\,.
\eeq
For generic $\alpha$, there is no way to solve this equation: the scaling with $k_1$ is different on each sides.  We cannot resolve this problem including more terms in the soft limit, as $k_1^{\alpha}$ is the smallest power of $k_1$ by definition. The only resolution is therefore that each side vanishes independently, which means
\beq
\alpha (\alpha-1) + 4 \alpha = 0 \then \alpha = -3, 0 \,.
\eeq 
Clearly $\alpha =-3$ violates our assumption that the single field consistency condition are valid\footnote{Note that the complete solution in \eqref{eq:conformal_bispectrum} also contains a $k_1^{-3} \log K$ term which still behaves as $\alpha=-3$ in the soft-limit.}, \eqref{eq:sfcc}.  The case $\alpha = 0$ requires more discussion as it is consistent with \ref{eq:sfcc}.  

The case $\alpha = 0$ corresponds to a function that is purely analytic in $\k_1$.  Certainly $k_1^{\alpha=0}$ is analytic. Then, \eqref{eq:bi_constraint} relates the coefficient of $k^{\alpha -2}$ to terms of order $k^\alpha$.  Therefore, the coefficient of $k_1^0$ fixes the soft limit of the all the $k^{2n}$ terms where $n$ is a positive integer.  This entire series is analytic in $\vec k_1$ and thus is the Fourier transform of a $\delta$-function and derivatives thereof (i.e.~contact terms).  Furthermore, inverting the logic, introducing a term that was non-analytic would required an $\alpha =-3$ term (or an infinite series of arbitrarily negative powers of $k_1$).   The corresponding three-point correlation function (i.e.~position space) is therefore zero for some open set of $\x_1$.  We are restricting ourselves to correlation functions that are non-zero for generic $\vec x_i$ and thus we exclude $\alpha =0$ as a valid solution of this kind.  

Now let's repeat this argument for a generic $N$-point function of $\zeta$.  Unlike the bispectrum, higher point correlation functions can depend on the relative angles between momenta.  As a result, the soft limit generically takes the form
\beq
\lim_{k_1 \to 0} \langle \zeta_{\k_1} \prod_{a=2}^N \zeta_{\k_a} \rangle'  = k_1^{\alpha} \prod_{a=1}^m (\k_1\cdot \vec P_a(\{\vec k_{b >1}\}) ) \times F(\{\vec k_{b >1}\})\,,
\eeq
where $ \vec P_a(\{\vec k_{b >1}\})$ is some linear combination of $\vec k_{b}$ for $  b=2,3\dots,N $, and $m \geq 0$ is an unknown integer. We again require that the special conformal transformations annihilate the correlation function, $\vec b \cdot \vec K\langle \zeta_{\k_1} \prod_{a=2}^N \zeta_{\k_a} \rangle'  = 0$.  However, for a generic $N$-point function there is no simplification of the differential operator, and therefore we require that
\beq
\sum_{a=1}^N\left[6 \vec{b} \cdot \vec{\partial}_{k_{a}}-\vec{b} \cdot \vec{k}_{a} \vec{\partial}_{k_{a}}^{2}+2 \vec{k}_{a} \cdot \vec{\partial}_{k_{a}}\left(\vec{b} \cdot \vec{\partial}_{k_{a}}\right)\right]\langle \zeta_{\k_1} \prod_{b=2}^N \zeta_{\k_{b}} \rangle'  = 0 \ .
\eeq
Taking the soft limit, we have
\beq
\left[6 \vec{b} \cdot \vec{\partial}_{k_{1}}-\vec{b} \cdot \vec{k}_{1} \vec{\partial}_{k_{1}}^{2}+2 \vec{k}_{1} \cdot \vec{\partial}_{k_{1}}\left(\vec{b} \cdot \vec{\partial}_{k_{1}}\right)\right] k_1^{\alpha} \prod_{a=1}^m (\k_1\cdot \vec P_a) F(\{\vec k_{b >1}\}) = k_1^{\alpha} \prod_{a=1}^m (\k_1\cdot \vec P'_a)  G(\vec b, \{\vec k_{b >1}\})\,,\nonumber
\eeq
where 
\beq
\prod_a \vec P'_a  G(\vec b, \{\vec k_{b >1}\}) = \vec b \cdot \vec K \prod_a  \vec P_a  F( \{\vec k_{b >1}\})\,.
\eeq
Applying the differential operator, one finds
\bea
&& \left[6 \vec{b} \cdot \vec{\partial}_{k_{1}}-\vec{b} \cdot \vec{k}_{1} \vec{\partial}_{k_{1}}^{2}+2 \vec{k}_{1} \cdot \vec{\partial}_{k_{1}}\left(\vec{b} \cdot \vec{\partial}_{k_{1}}\right)\right] k_1^{\alpha} \prod_{a=1}^m (\k_1\cdot \vec P_a) \\
&=& 6\left(\alpha \b\cdot \k_1 k_1^{\alpha -2} \prod_{a=1}^m(\k_1\cdot \vec P_a) + k_1^{\alpha} \sum_{b=1}^m \b \cdot \vec P_{b} \prod_{a \neq b} (\k_1\cdot \vec P_a)\right) \nonumber \\
&& - \b\cdot \k\left( (\alpha (\alpha+1) + 2\alpha m)k^{\alpha-2}\prod_{a=1}^m  (\k_1\cdot \vec P_a) +k_1^\alpha \sum_{b\neq c=1}^m \vec P_c \cdot \vec P_{b} \prod_{a \neq b,c} (\k_1\cdot \vec P_a)\right) \\
&& + 2 \Bigg(  \alpha (\alpha -1+m)  \, (\b\cdot \k_1) k_1^{\alpha-2} \prod_{a=1}^m  (\k_1\cdot \vec P_a)  + (\alpha+m-1) k_1^{\alpha}  \sum_{b=1}^m \b \cdot \vec P_b \prod_{a \neq b} (\k_1\cdot \vec P_a) \Bigg) \nonumber \\
&=&k_1^{\alpha -2} \b\cdot \k_1  \prod_{a=1}^m(\k_1\cdot \vec P_a)  \Bigg(    6 \alpha - \alpha (\alpha+1) + 2 \alpha (\alpha -1)  \Bigg)  \\
&&+k_1^{\alpha} \sum_{b=1}^m \b \cdot \vec P_b \prod_{a \neq b} (\k_1\cdot \vec P_a) \Bigg(6 +2 \alpha +2 (m-1) \Bigg) \label{eq:pcdotb} \\ && - \b\cdot \k_1 k_1^\alpha \sum_{b\neq c=1}^m \vec P_c \cdot \vec P_b \prod_{a \neq b,c} (\k_1\cdot \vec P_a) \label{eq:pipj}\,.
\eea
The final term, (\ref{eq:pipj}), has no analogue on the right-hand side of this equation and does not vanish on the left-hand side for $m >1$.  Therefore, there can be no solution with $m>1$ and we can reduce this problem to $m=0$ and $m=1$.  For $m=0$, the second to last line, (\ref{eq:pcdotb}), also vanishes and we have $\alpha = -3,0$ as solutions\footnote{One might be concerned that we are missing the possibility of $k_1^{-3} \log k_1$  or $\log k_1$ terms.  The appearance of such logs can be understood as the subleading term from the limit $k^{\alpha}$ as $\alpha \to -3$ or $0$ and will not alter any conditions regarding the consistency conditions.}. Just like for the bispectrum, $  \alpha=-3 $ violated the squeezed limit consistency relation, while $  \alpha=0 $ does not produce a non-zero correlation function for generic $\x_i$ (i.e.~$\alpha=0$ gives a series of contact terms). For $m=1$, the only possible solution is $\alpha =-3$ which means the unique behavior in the soft limit is
\beq
\lim_{k_1 \to 0}  \langle \zeta_{\k_1} \prod_{a=2}^N \zeta_{\k_a} \rangle' \propto \frac{\vec k_1 \cdot \vec P}{k_1^3}\,.
\eeq
For example, this is precisely the soft behavior found in the trispectrum of a massless scalar with derivative interactions~\cite{Creminelli:2011mw}.  We will now show that $m=1$ also violates the single field consistency conditions.  If we apply the non-linearly realized conformal transformation, $\vec K_{\rm NL}$ then we have
\beq
\lim_{k_1 \to 0} \left\langle\zeta_{\vec{k}_1} \zeta_{\vec{k}_{2}} \cdots \zeta_{\vec{k}_{N}}\right\rangle=-\frac{1}{2} P(k_1) \vec k_1 \sum_{a=2}^{N}\left(6 \partial_{k_{a}}^{i}-k_{a}^{i} \vec{\partial}_{k_{a}}^{2}+2 \vec{k}_{a} \cdot \vec{\partial}_{k_{a}} \partial_{k_{a}}^{i}\right)\left\langle\zeta_{\vec{k}_{2}} \cdots \zeta_{\vec{k}_{N}}\right\rangle^{\prime}\,.
\eeq
Note that the right-hand side is just the conformal transformation of the $(N-1)$-point function. Invariance under $  K $ requires that the differential operator annihilates the correlator and therefore
\beq
\sum_{a=1}^{N}\left(6 \partial_{k_{a}}^{i}-k_{a}^{i} \vec{\partial}_{k_{a}}^{2}+2 \vec{k}_{a} \cdot \vec{\partial}_{k_{a}} \partial_{k_{a}}^{i}\right)\left\langle\zeta_{\vec{k}_{1}} \cdots \zeta_{\vec{k}_{n}}\right\rangle^{\prime} = 0\,.
\eeq
This proves that also the $m=1$ solution must vanish to respect the non-linearly realized special conformal transformation. We conclude that invariance under the set of symmetries in \eqref{syms}-\eqref{syms2} implies that all connected $  N $-point functions vanish for $  N \geq 3 $ and hence the theory is free.


\subsection{Second proof: the operator product expansion} \label{ope}

In this section, we will use the operator product expansion (OPE) to show that any connected $ N  $-point correlator of $  \zeta $ that is invariant under the symmetries in \eqref{syms}-\eqref{syms2} must vanish identically. In the first part of the proof we will show that $  \zeta $ cannot appear non-trivially in the OPE. In the second part, we will show that any other operator that is build out of massive fields of any spin also cannot appear.

By construction, the decoupling limit isolates the scalar metric mode, $\zeta(\x, t)$, from the tensors and the non-dynamical components of the metric (i.e.~the ADM components $N$, $N^i$).  In this limit, the action can be written as a local function of $\zeta(t,\x)$ (and other local fields), which ensures that $\zeta(\x)$ obeys an OPE.  We want to prove that this OPE is constrained by symmetry to be that of the free theory, namely\footnote{Here and in the following, all product of operators at the same point should be considered renormalized normal products and in particular normal ordered.} (we leave spatial indices implicit to avoid clutter)
\begin{align}
\zeta(\x)\zeta(0) \overset{!}{=} \left[ \sum_{a=0}^{\infty}\frac{\x^{n}}{n!}\vec \partial^{n}\zeta(0) \right]\zeta(0)\quad \text{(free theory)}\,.\label{free}
\end{align}
To see this, let's subtract from the full OPE the OPE of the free theory and define\footnote{We used translation invariance to set $  \y=0 $. In computing the variation of this operator one should be careful though that derivatives do not commute with the $  y\to 0 $ limit.}
\begin{align}\label{both}
\zeta(\x)\zeta(\y)- \left[ \sum_{a=0}^{\infty}\frac{\x^{n}}{n!}\vec \partial^{n}\zeta(0) \right]\zeta(0)=\sum_{n}c_{n}(\x)\O_{n}(0)\,,
\end{align}
where both the c-numbers $  c_{n}(\x) $ and the operators $  \O_{n} $ can in general have implicit spacial indices that are contracted with each other, such as for example $  c_{ij}\O^{ij}$. Now we will show that all the non-free terms in the OPE must vanish, $  c_{n}(\x)=0 $.  Recall that in Section \ref{ssec:algebra} we had found infinitely many symmetry generators, namely $  Q $, $  V_{i} $, $  V_{ij} $, $  V_{ijk} $ and their higher-order cousins $ V_{(n)}  $. It is the action of these symmetries on \eqref{both} that sets all the $  c_{n}(\x) $ to zero. To see this, let's start acting with $  Q $ on both sides of \eqref{both}. On the left-hand side one finds
\begin{align}
[Q,\zeta(\x)\zeta(0)- \left[ \sum_{a=0}^{\infty}\frac{\x^{n}}{n!}\vec\partial^{n}\zeta(0) \right]\zeta(0)]&=\zeta(\x)+\zeta(0)-\left[ \sum_{a=0}^{\infty}\frac{\x^{n}}{n!}\vec \partial^{n}\zeta(0) \right]-\zeta(0)=0\,.
\end{align}
So acting with $  Q $ on the right-hand side of \eqref{both} must also give zero
\begin{align}
\left[ Q,\sum_{n}c_{n}(\x)\O_{n}(0) \right] \overset{!}{=}0\,.
\end{align} 
But this means that $  \O_{n} $ can contain only products of $  \zeta $ with at least one derivative acting on each $  \zeta $. In particular, all operators of the form $  \O_{n}\sim \zeta \tilde \O_{n-1} $ for any $  \tilde \O_{n-1} $ cannot appear (i.e.~the respective $  c_{n} $ must vanish). We can then act with $  V_{i} $ on both sides of \eqref{both}. Again the left-hand side vanishes:
\begin{align}
[V_{i},\zeta(\x)\zeta(0)- \left[ \sum_{a=0}^{\infty}\frac{\x^{n}}{n!}\vec \partial^{n}\zeta(0) \right]\zeta(0)]=0\,.
\end{align}
Requiring that the right-hand side vanishes as well, we find that $  \zeta $ can appear on the  right-hand side of the OPE only with two derivatives acting on it, namely all the coefficients $  c_{n} $ of operators of the form $  \O_{n}\sim \partial_{i}\zeta \tilde\O_{n-1} $ must vanish for any $ \tilde\O_{n-1}  $. Now we act with $  V_{li} $ on both sides of \eqref{both}. Again, the left-hand side vanishes
\begin{align}
[V_{li},\zeta(\x)\zeta(0)- \left[ \sum_{a=0}^{\infty}\frac{\x^{n}}{n!}\vec \partial^{n}\zeta(0) \right]\zeta(0)]=0\,,
\end{align}
and so the right-hand side has to vanish as well. To see that this forbids all operators $  \partial_{ij}\zeta $ with two derivatives acting on $  \zeta $ from appearing on the right-hand side, let see how $  \partial_{ij}\zeta $ transforms under $  V_{lm} $,
\begin{align}
[V_{lm},\partial_{ij}\zeta]=\delta_{li}\delta_{mj}+\delta_{lj}\delta_{mi}-\delta_{lm}\delta_{ij}\,.
\end{align}
The most generic operator containing $  \partial_{ij}\zeta $ in the OPE must be of the form
\begin{align}
c_{n}(\x)\O(0)=\partial_{ij}\zeta(0) F_{ij}\,,
\end{align}
for some function $  F_{ij} $ of the field $  \zeta(0) $, its derivatives and the coordinate $  \x $, where only the symmetric part of $  F^{ij} $ contributes. Using the commutator above one finds
\begin{align}
0=[V_{lm},c_{n}(\x)\O(0)]=2 F_{lm}-\delta_{lm} F_{ii}\,.
\end{align}
The trace of this expression implies $  F_{ii}=0 $, from which $  F_{lm}=0 $ follows. 
One can continue using the higher order symmetries $  V_{(n)}\equiv V_{i_{1}\dots i_{n}} $ to show that the $  n $-th derivative of $  \zeta $ cannot appear on the right-hand side of the OPE for any $  n $. To convince oneself that this procedure sets to zero operators containing $  \zeta $ with any number $  n $ of derivatives, it suffices to show that the number of contraints matches the number of free coefficients. Let's define
\begin{align}
c_{n}(\x)\O(0)=F_{i_{1}\dots i_{n}} \partial_{i_{1}\dots i_{n}}\zeta \,.
\end{align}
Since $  F_{i_{1}\dots i_{n}} $ is totally symmetric it has $  (2+n)(1+n)/2 $ components. But the symmetry generator $ V_{(n)} $ also has precisely the same number of components, each of which constrains a different linear combination of the coefficients of $  F_{i_{1}\dots i_{n}} $, which therefore must all vanish. We conclude that $  \zeta $ with any number of derivatives cannot appear on the right-hand side of the OPE in \eqref{both}.


What about operators that do not contain $  \zeta $? We will now show that also all operators $  \O_{n} $ of non-vanishing conformal dimension, $\Delta_n >0$, are forbidden by symmetry to appear in the OPE in \eqref{both}. Indeed we can already anticipate that we should not be able to forbid operators of zero dimension from appearing in the OPE. These would correspond to massless fields and we know that in the presence of additional massless scalar fields, the consistency relations for $  \zeta $ do not apply. This in turn means we cannot take advantage of $  D_{\text{NL}} $ and $  K_{\text{NL}} $, which were crucial in the proof we gave in the previous subsection.

Since $  \Delta_{\zeta}=0 \neq \Delta_{n} $, applying the special conformal transformations to the two-point function implies
\beq\label{difdim}
\langle \zeta(\x) \O_n(0) \rangle = 0\,,
\eeq
On general grounds, anytime an operator $\O$ appears in the OPE of operators $\phi_1$ and $\phi_2$ we can construct a non-zero three point function of the form $\langle \phi_1 \phi_2 \O\rangle$. In fact, the special conformal transforms completely fix the form of this three-point function up to a constant. Consider now the case in which $\phi_1$ and $\phi_2$ have spin-zero and $\mathcal{O}_{3}^{i_{1} \cdots i_{\ell}}$ has spin $\ell$, then by conformal symmetry the three-point function must take the form~\cite{Costa:2011mg,Simmons-Duffin:2016gjk}
\beq
\left\langle\phi_{1}(x_{1}) \, \phi_{2}(x_{2}) \, \mathcal{O}_{3}^{i_{1} \cdots i_{\ell}}(x_{3})\right\rangle=\frac{f_{123}\left(\widehat{Z}_{3}^{i_{1}} \cdots \widehat{Z}_{3}^{i_{\ell}}-\text { traces }\right)}{x_{12}^{\Delta_{1}+\Delta_{2}-\Delta_{3}} x_{23}^{\Delta_{2}+\Delta_{3}-\Delta_{1}} x_{31}^{\Delta_{3}+\Delta_{1}-\Delta_{2}}}\,,
\eeq
where
\begin{align}
x_{ab}&\equiv \sqrt{(\x_{a}-\x_{b})\cdot (\x_{a}-\x_{b})}\,, & Z_{3}^{i} &\equiv \frac{x_{13}^{i}}{x_{13}^{2}}-\frac{x_{23}^{i}}{x_{23}^{2}},  &\widehat{Z}_{3}^{i}&=\frac{Z_{3}^{i}}{\left|Z_{3}\right|} \,,
\end{align}
and $f_{123}$ is a free coefficient, which is nothing but the coefficient of $  \O $ in the OPE of $  \phi_{1} $ and $  \phi_{2} $.  Applying this to the case of interest we take $\phi_1 = \phi_2 =\zeta$ with $\Delta_{1}= \Delta_2 = 0$.  The important consequence of this formula is that it constrains the form of the OPE of $\zeta$ and $\mathcal{O}_{3}^{i_{1} \cdots i_{\ell}}$.  Specifically, since $\langle \zeta(\x) \O_n(0) \rangle = 0$ for generic operators, the limit 
\beq
\lim_{\x_2 \to \x_3} \left\langle \zeta (\vec x_{1}) \, \zeta(\vec x_{2}) \, \mathcal{O}_{3}^{i_{1} \cdots i_{\ell}}(\vec x_{3})\right\rangle  = f_{\zeta \zeta \O} \frac{(-1)^{\ell} x_{23}^{i_1} \dots x_{23}^{i_\ell}}{x_{23}^{\Delta_3 + \ell}} +\dots
\eeq 
isolates the $\zeta$ piece of the OPE\footnote{Since $  \Delta_{\zeta}=0 $, $  \ex{\zeta(\x)\zeta(\y)} $ is just a constant, independent of $  \x $ and $\y $, which is fixed by the amplitude of the scalar power spectrum.  Technically speaking, $  \ex{\zeta(\x)\zeta(\y)} $ should also include $\log(|\x-\y|)$ from the sub-leading term in the $\Delta_\zeta \to 0$ limit.  This detail is irrelevant for the purpose of our discussion.}
\beq
\lim_{\vec x\to 0}\zeta(\x) \O^{i_{1} \cdots i_{\ell}}(0) \supset \frac{f_{\zeta \zeta \O}}{ \langle \zeta(\x)\zeta(0)\rangle}  \ \frac{(-1)^{\ell} x^{i_1} \dots x^{i_\ell}}{x^{\Delta_3 + \ell}}  \zeta(0) \ .
\eeq
The fact that $\zeta(0)$ appears on the right-hand side suggests that $f_{\zeta \zeta \O}$  will be strongly constrained by the nonlinearly realized symmetry. Indeed, if we apply $D_{\rm NL}$ to both sides of the equation, we get
\beq
\x \cdot \partial_\x (\zeta(\x) \O^{i_{1} \cdots i_{\ell}}(0) ) + \zeta(\x) ( D_{\rm NL} \O^{i_{1} \cdots i_{\ell}}(0) )  + \O^{i_{1} \cdots i_{\ell}}(0)  \supset  \frac{f_{\zeta \zeta \O}}{ \langle  \zeta(\x)\zeta(0)\rangle}  \ \frac{(-1)^{\ell} x^{i_1} \dots x^{i_\ell}}{x^{\Delta_3 + \ell}} \,.
\eeq 
The expectation value of the left-hand side vanishes because of \eqref{difdim}. So the expectation value of the right-hand side must also vanish and we conclude that $  f_{\zeta\zeta\O}=0 $. This shows that no operator of dimension $  \Delta \neq 0 $ can appear on the right-hand side of \eqref{both}.

Summarizing, we have shown that the OPE of two $  \zeta $'s must be that of a free theory, which in real space takes the form of \eqref{free}. We can use this OPE to compute any $  n$-point connected correlator. The fact that the real-space OPE limit in \eqref{free} is analytic around $  x=0 $ ensures that the correlators in momentum space, obtained by the Fourier transform, must decay exponentially for large momenta. But by virtue of scale invariance, correlators must scale as power laws in the momenta and so the only possibility is that all connected correlators must vanish, hence proving Theorem 2.


\subsection{Including dynamical gravity}

Our analysis thus far has been restricted to the slow-roll decoupling limit where $\Mpl \to \infty$ but the scalar power spectrum remains constant.  This was a necessary condition for producing exactly scale-invariant scalar correlators.  In addition, it means that the tensor fluctuations $\gamma_{ij}$ decouple.  In this sense, our scalar fluctuations are described by a local QFT on a fixed de Sitter background.

A natural question is what happens at finite $\Mpl$.  We certainly expect a free quantum field theory coupled to dynamical gravity will be non-Gaussian. Very naively, one might expect that the size of the non-Gaussian contributions to the $\zeta$-correlations would be bounded by the first slow-roll parameter, $  \e $, which controls deviations from de Sitter and hence exact scale invariance, i.e.~non-Gaussianity would be ${\cal O}(\epsilon)$.  Specifically, in the limit, $\epsilon \to 0$, the geometry might seem to asymptote pure de Sitter and, for example, we would expect the tensors correlators to be invariant under linearly realized conformal transformations~\cite{Maldacena:2011nz}. However, this is not quite true when considering correlators involving $\zeta$. For example, the $  \zeta $ bispectrum has physical contributions of $  \O(\eta) $, without any $  \e $ suppression \cite{Cabass:2016cgp}. As another example, the scalar trispectrum due to graviton exchange is not Gaussian even in the $\epsilon,\eta \to 0$ limit~\cite{Ghosh:2014kba}.  

The origin of the unexpected contribution to the trispectrum is that $\zeta$ is not just a spectator field in a de Sitter geometry\footnote{See e.g.~the discussion in Section 5.1 of \cite{Cabass:2016cgp}}.  Instead, $\zeta$ is a component of the metric and breaks the isometry we would associate with de Sitter boosts $K^{i}$.  Furthermore, the tensor modes do not respect the nonlinear transformation, $K_{\rm NL}$, while leaving the gauge fixed~\cite{Hinterbichler:2013dpa,Ghosh:2014kba}.  The net result is that the presence of dynamical tensor modes alone will break the conformal symmetry, even in the limit when the background is pure de Sitter. This is seen most clearly in the scalar-scalar-tensor bispectrum~\cite{Maldacena:2002vr, Mata:2012bx}
\beq
\left\langle\gamma_{\vec{k}_{1}}^{s} \zeta_{\vec{k}_{2}} \zeta_{\vec{k}_{3}}\right\rangle'= \Delta_\gamma^2 \Delta_\zeta^2 \frac{4}{\prod\left(2 k_{i}^{3}\right)}  \epsilon_{i j}^{s} k_{2}^{i} k_{3}^{j}  \left( -k_{t}+\frac{\sum_{i>j} k_{i} k_{j}}{k_{t}}+\frac{k_{1} k_{2} k_{3}}{k_{t}^{2}} \right)\,,
\eeq
where $k_t = k_1+k_2+k_3$. Here $\Delta_\zeta$ and $\Delta_\gamma$ are the amplitudes of scalar and tensor modes respectively.  This result is consistent with $\gamma_{ij}$ appearing in the OPE of  $\zeta$ with an order one coefficient.  Repeating the OPE argument, this is not consistent with linearly realized conformal invariance and so there is no contradiction with our results.  It was shown in~\cite{Ghosh:2014kba} that the correlation functions of $\zeta$ will still obey a non-linearly realized symmetry related to the additional gauge transformation required to maintain the gauge of $\gamma_{ij}$ after performing the $K_{\rm NL}$ transformation.  

However, it is important in this argument that we are imposing the $K_{\rm NL}$ on the OPE and not purely as constraint on the scalar-scalar-tensor bispectrum.  Indeed, as shown in~\cite{ Mata:2012bx}, one can use approximate conformal invariance to determine this particular correlation function but fails in higher point correlators~\cite{Ghosh:2014kba}.  This phenomena is not limited to tensor fluctuations.  For example, It was observed in~\cite{alberto} that assigning a non-zero scaling dimension to $\zeta$ can be useful in determining the expression for the spectral tilt, $n_s-1$.  This works because the quadratic action possess an approximate scaling symmetry, even though such a symmetry is never exact.  Similarly, approximate (linearly realized) conformal invariance has been used as an intermediate step in deriving the bispectrum in (quasi) single field inflation~\cite{Arkani-Hamed:2018kmz}, even though the resulting expression is not the one fixed by conformal invariance.  In this sense, linearly realized conformal invariance may be useful in determining the structure of specific correlation functions, particularly with gravitational strength interactions, even though, as we proved here, it cannot hold as an exact symmetry of the full theory.  

To summarize, the coupling to dynamical gravity does not violate our theorem.  The coupling between the scalar and tensor modes introduces non-Gaussian correlations but also breaks conformal symmetry, even in the limit where the geometry is well approximated by de Sitter space.  The contributions of tensors to the non-Gaussian correlators of $\zeta$ are well below the sensitivity of cosmological surveys in the conceivable future.  Most non-Gaussian correlators at current levels of sensitivity are well approximately by the $M_{\rm pl} \to \infty$ limit of the effective action for the fluctuations, and thus the question of whether they can obey linearly realized conformal symmetry (or other symmetries) is still observationally relevant.


\subsection{Holographic interpretation}\label{holog}

Holography provides a natural framework to understand the constraints of spacetime symmetries on de Sitter correlators. In pure de Sitter space, the isometries of the background imply the wave-function of the universe is the partition function of a non-unitary conformal field theory (CFT)~\cite{Witten:2001kn,Strominger:2001pn,Mazur:2001aa,Maldacena:2002vr,Maldacena:2011nz}.   The behavior of fluctuations in an inflationary spacetime has a natural interpretation in terms of the RG flow in a holographic dual~\cite{Henningson:1998gx,Bianchi:2001de,Bianchi:2001kw,Strominger:2001gp,McFadden:2009fg}.  In this description, $\zeta$ is dual to the trace of the stress tensor of the dual theory.

Naively, it would seem that we have demonstrated a well-known result in QFT, namely that the trace of the stress tensor vanishes at a conformal fixed point.  This is not the correct interpretation of our result, as the trace of the stress tensor does not vanish in the decoupling limit~\cite{Baumann:2019ghk}.  By construction, we are holding the power spectrum of $\zeta$ fixed with amplitude $\Delta_\zeta^2$.  The holographic interpretation of the wavefunction then determines
\beq
\langle \zeta_\k \zeta_{\k'} \rangle' = \frac{2 \pi^2 \Delta_\zeta^2}{k^3} = \frac{1}{-2 {\rm Re}\langle T_\k T_{\k'} \rangle' }\,,
\eeq
where $T \equiv T_i^i$ is the trace of the stress tensor.  Since $\Delta_\zeta \ll 1$, not only is $T(\x) \neq 0$ but its power spectrum is very large (although still small compared to the central charge which diverges in the decoupling limit).  Adding higher derivative interactions to the bulk allows a wide variety of non-trivial correlation functions of $T$ while preserving scale (but not conformal) invariance.

The origin of the confusion is precisely the $\Mp \to \infty$ limit.  At any finite $\Mpl$, for a generic model of inflation, scale invariance is broken at order $\epsilon$ while conformal invariance is broken at ${\cal O}(1)$~\cite{Baumann:2019ghk}.  As a result, generic interacting inflationary models are approximately scale but not conformal, but neither is exact (they only become exact in the decoupling limit).  While scale-but-not-conformal behavior is generally not expected in the QFT~\footnote{Technically speaking, the QFT duals in dS/CFT are non-unitarity and thus are not subject to the usual QFT intuition.  However, as explained in~\cite{Baumann:2019ghk}, an essentially identical pattern of symmetries appear in AdS where the QFT intuition is applicable.}, the breaking of scale at order $\epsilon$ ensures it is consistent with known results~\cite{Luty:2012ww}.

Surprisingly, the simple pattern of symmetry breaking described in the introduction does not have a simple holographic interpretation.  The reason is that the bulk description relies on the presence of an approximate global symmetry which is only an exact symmetry in the $\Mpl \to \infty$ limit.   While such a symmetry is well motivated in the bulk EFT, it cannot be a fundamental symmetry in a theory of quantum gravity and therefore is not manifest in the CFT.  This can be seen already with the approximate shift symmetry for a scalar field, which would be dual to a family of approximately scale-but-not-conformally invariant field theories, related by a marginal deformation, but where correlation functions are unchanged by the marginal deformation~\cite{Harlow:2018tng}.  This behavior is not found in known QFTs but must arise in interacting theories to be compatible with known mechanisms of inflation~\cite{Baumann:2019ghk}.  The interpretation of Theorem 2 is that if there is a family of approximately conformally invariant theories with non-vanishing $T$ (rather than just scale invariant), then every theory in this family is free. To our knowledge, there is no proof of this statement in QFT.

 
\section{Theorem 3: all linearly-realized symmetries of a single scalar}\label{sec:linear}

In this section we derive a complete classification of all the possible linearly-realized symmetries of the correlators of a massless scalar field $  \phi $ in an accelerated FLRW universe, namely all the possible relations of the form
\begin{align}\label{obey}
\sum_{a=1}^{n} f(\vec k_{a},\partial_{\vec k_{a}}) \ex{\phi_{ \vec k_{1}} \dots \phi_{\vec k_{n}}}=0
\end{align} 
that can be satisfied by a non-vanishing set of correlators for all $  n $ and some non-vanishing continuous functions $  f $. The assumptions of the theorem are the following:
\begin{enumerate}
\item The correlators are statistically homogeneous, isotropic and scale invariant, i.e.~they are invariant under $  P_{i} $, $  M_{ij} $ and $  D $ in \eqref{syms}-\eqref{3b}.
\item The time dependence of correlators in the asymptotic future can be neglected.
\item The symmetry transformations are local in space in the sense that they do not involve inverse Laplacians.
\item There is a finite number of times that a generator $  Q $ can be commuted with space translations $  P_{i} $ leaving a non-zero result. In formulae, there exist some finite $  N $ for which
\begin{align}\label{ass}
[P_{i_{N}}[P_{i_{N-1}} \dots [P_{i_{1}},Q]]=0\,.
\end{align}
\end{enumerate}
In this section, we make no assumptions about the particle content of the theory, the nature of interactions or the slow-roll and decoupling limit. Under the assumptions above, we will prove that \textit{the only additional linearly-realized symmetries that the correlators of a single scalar field can satisfy, besides rotations, translations and dilations, are special conformal transformations, $  K^{i} $}. By combining this theorem with Theorem 2, we can obtain an interesting lemma. In the particular case in which the scalar field in the above theorem is $ \zeta $, namely curvature perturbations, the symmetry algebra must also include $  D_{\text{NL}} $ and $  K_{\text{NL}} $. Then we know that the addition of $  K^{i} $ to the symmetry algebra enforces the theory to be free by virtue of Theorem 2. Therefore, for $  \zeta $ correlators with the soft limits dictated by the single-field consistency relations, the largest possible set of linearly-realized symmetries is composed precisely by those symmetries that we have already observed in the primordial power spectrum, namely homogeneity, isotropy and scale invariance, with dilations being only an approximate symmetry. In other words, assuming single field inflation, the primordial perturbations in our universe display the largest possible amount of symmetry for an interacting theory. This is very reminiscent of the situation in flat space, where the Coleman-Mandula theorem ensures that the largest possible set of linearly-realized spacetime symmetries is just the Poincar\'e group (assuming the S-matrix is well defined).

The theorem in this section can also be rephrased as follows. Assuming a set of correlators that obey \eqref{obey} for 
\begin{align}
P^{i}&: f^{i}(k,\partial_{k})=-ik^{i} & &\text{(translations)}\,, \\
M_{ij}&:f^{i}_{j}(k,\partial_{k})=2k^{[i}\partial_{k_{j]}} & &\text{(rotations)}\,, \\
D&:f(k,\partial_{k})=3-\Delta + \vec k\partial_{\vec k} & &\text{(dilations)}\,,
\end{align}
where $  \Delta $ is the scaling dimension of $  \phi $, the only additional function $  f $ that can satisfy \eqref{obey} for all $  n $ is 
\begin{align}
f_{i}(k,\partial_{k})=i\left[ 2 \vec{k} \cdot \partial_{\vec k} \partial_{k^{i}}-k_{i}\partial_{\vec k}\cdot \partial_{\vec k}+2\left(  3-\Delta\right) \partial_{k^{i}} \right]\,.
\end{align}

As in the proof of the Coleman-Mandula and Haag-Lopuszanski-Sohnius theorems \cite{Coleman:1967ad,Haag:1974qh}, or more recently in \cite{Pajer:2018egx,Grall:2019qof}, it is useful to organize the proof in terms for the degree $ N $ of a symmetry generator $  S^{N} $, which is defined as the largest power of $ x $ that appears in the associated symmetry transformation
\begin{equation}
\delta_{S^N_m}\phi=\sum_{n=0}^N\,x^{i_{1}}\,\dots\,x^{i_n}\,f_{i_1 \dotsi_ni_{n+1}\,\dots\,i_{n+m}}\,,
\end{equation}
where the lower index $ m$ counts the number of spacial indices\footnote{It would be more natural to talk about irreps of SO$(3) $ rather than number of indices, hence separating trace, symmetric-traceless, anti-symmetric parts. Instead of setting up this notation, we will take advantage of this separation only when needed in the proof.} and $  f $ is a \textit{linear} function of $  \phi $ and its derivatives. This is useful because the commutator of $  S^{N}_{m} $ with spatial translations gives
\begin{align}
[S^{N}_m,P^i]=S^{N-1}_{m+1}\,.
\end{align}
The strategy is to show that there are no degree zero generators beside $  P_{i} $ and no degree one generators beside $  M_{ij} $ and $  D $. Higher degree generators are then constrained by the above relation in that they have to reduce to $  P_{i} $, $  M_{ij} $ and $  D $ upon commuting enough times with translations.

 
\subsection{Degree zero}

Let's start with generators of degree zero with any number of spacial indices, $ S^{0}_{m}$. When acting on the field they take the form
\begin{align}
 S^{0}_{m}&: \delta \phi(\vec k)=f_{i_{1}\dots i_{m}}(\vec k) \phi(\vec k)\,,
\end{align} 
for some non-vanishing $  f_{i_{1}\dots i_{n}}(\vec k) $, whose indices come from $  k_{i} $, $  \delta_{ij} $ or $  \e_{ijk} $. Demanding invariance of an $ n$-point correlator under $ S^{0}_{m}$ implies
\begin{align}
\sum_{a=1}^{n} f_{i_{1}\dots i_{n}}(\vec k_{a})  \ex{\phi( \vec k_{1})\dots \phi(\vec k_{n})}=\sum_{a=1}^{n} f_{i_{1}\dots i_{n}}(\vec k_{a}) B_{n}\delta_{D}^{3}\left(  \sum_{b=1}^{n}\vec k_{b}\right)&=0\,.
\end{align} 
Stripping away the momentum-conserving delta function we find
\begin{align}\label{sat}
B_{n}\sum_{a=1}^{n} f_{i_{1}\dots i_{n}}(\vec k_{a}) =0\,,
\end{align}
where $  \vec k_{n} =-\sum_{a}^{n-1}\vec k_{a} $. We want to prove that any $  B_{n} $ that satisfies this relation must vanish for \textit{generic} kinematic input\footnote{This is much easier to prove than the more general statement that the correlators must vanish everywhere, and it is also what is proven by the Coleman-Mandula theorem or the Weinberg's soft theorems. It seems likely that correlators that are non-vanishing only on some codimension one or larger hypersurface in momentum space will violate cluster decomposition, but we have not proven this in detail.}. We will prove this by contradiction. Suppose there was a non-vanishing set of $  B_{n} $ that satisfies \eqref{sat} for generic momenta. Then we would conclude 
\begin{align}
\sum_{a=1}^{n} f_{i_{1}\dots i_{n}}(\vec k_a) = 0\,,
\end{align}
for generic momenta. But this can only happen if $  f_{i_{1}\dots i_{n}}(\vec k) $ vanishes identically, contradicting our assumptions. 

 
\subsection{Degree one}

The most generic symmetry of degree one must take the form\footnote{One could also start with the real space action of the symmetry. But usually we work with correlators in Fourier space and it is more convenient to work with symmetry transformations in Fourier space as well. Notice that the Fourier transform of the transformation gets a bit messy because, when integrating by part in $  \partial_{x} $ one hits the factors of $  x $. The calculation is doable if one assumes some simple form of $  f $, e.g.~a polynomial. Anyways, starting directly in momentum space bypasses all these complications.}
\begin{align}
S^{1}_{m}: \delta \phi(\k)= \left[ f_{i_{1}\dots i_{m-1}}(\k) \partial_{ i_{m}}+ f_{i_{1} \dots i_{m}}(\k)\right]\phi(\k)\,,
\end{align}
where the spatial indices in the two $  f $'s can be taken by $  k_{i} $, $  \delta_{ij} $ or $  \e^{ijk} $.
The commutator with spatial translations tells us that
\begin{align}
[P_{i}, S^{1}_{m}]=S^{0}_{m+1}+\delta_{ij }S^{0}_{m-1}\,.
\end{align}
But we have already classified all possible degree zero symmetries and found that the only option are spatial translations, which have $ m=1$. Therefore, closure of the algebra demands that any new degree-one symmetry comes with zero or two spatial indices, $  m=2 $ or $  m=0 $, and hence takes the form 
\begin{align}\label{delta}
S^{1}_{2}: \delta \phi=f^{A}(k)k_{[i}\partial_{j]}+f^{ST}(k)k_{<i}\partial_{j>}+f^{LOT}(k)k_{<i}k_{j>}+\delta_{ij}\left[ f^{T}(k)  k^{l}\partial_{l}+f^{LO\delta}(k) \right]\,.
\end{align}
Here the indices stand for Anti-symmetric, Symmetric-Traceless, Lower Order Traceless, Trace and Lower-Order $  \delta $, respectively and $ <\dots>$ takes the symmetric traceless part,
\begin{align}
A_{<ij>}\equiv \frac{1}{2}\left(  A_{ij}+A_{ji}\right)-\frac{1}{3}\delta_{ij}A_{ll}\,.
\end{align}
Notice that the anti-symmetric, symmetric-traceless and traceless parts have to be independent symmetries as no cancellations among them are allowed by the index structure.
Let's explicitly calculate the following commutator (notice that $ [f(k),P_{i}]=0$ and so the functions of momenta can be treated as commuting factors)
\begin{align}
[k_{i}\partial_{j},P_{l}]=[k_{i}\partial_{j},k_{l}]=ik_{i}\delta_{jl}\,.
\end{align}
Since the commutator must reduce to a translation, which is the only degree zero symmetry, we conclude that $ f^{A}=f^{ST}=f^{T}=$ constant, while $ f^{LO\delta}$ and $  f^{LOT} $ remain unconstrained. In the anti-symmetric term we recognize rotations. 
So the only new symmetries can come from the symmetric traceless or trace transformations in \eqref{delta}. In general, for those transformations to be symmetries we need to demand
\begin{align}
\sum_{a=1}^{n}\left[  \vec k_{a} \vec \partial_{a} +f^{LO\delta}(k_{a}) \right] \ex{\phi^{n}}=0\,,\\
\sum_{a=1}^{n}\left[  k^{<i}_{a} \partial_{k^{j>}_{a}} +  f^{LOT}(k_{a})k^{<i}_{a}k^{j>}_{a} \right] \ex{\phi^{n}}=0\,.
\end{align}
Now we have the choice to work with operators that act on the full correlator, $ \ex{\dots}$, or on $ \ex{\dots}'=B_{n}$, where the delta function has been removed. For transformations of degree zero it is the same as the transformation acts multiplicatively. But for transformations of degree one or higher, one has to keep track of when $  \partial_{k} $ acts on the momentum-conserving delta function. Here is what happens for the two operators we care about
\begin{align}
\sum_{a=1}^{n} k^{i}_{a}\partial_{k_{a}^{j}} \delta^{3}_{D}\left(  \sum_{b}\vec k\right)&=\sum_{a=1}^{n} k^{i}_{a}\partial_{k_{a}^{j}} \int_{\vec x} e^{-i\v{x} \sum_{b} \v{k}}= \int_{\vec x}  \sum_{a=1}^{n} k^{i}_{a} (-i x^{j}) e^{-i\v{x} \sum_{b} \v{k}}\\
&= \int_{\vec x}   (-i x^{j}) \sum_{a=1}^{n} k^{i}_{a} e^{-i\v{x} \sum_{b} \v{k}}= \int_{\vec x}   (-i x^{j}) i \partial_{x^{i}} e^{-i\v{x} \sum_{b} \v{k}}\\
&= - \int_{\vec x}   \left( \partial_{x^{i}} x^{j} \right) e^{-i\v{x} \sum_{b} \v{k}}= - \delta_{ij} \delta^{3}_{D}\left(  \sum_{b}\vec k\right)\,.
\end{align}
So, when the trace transformations act on primed correlator there should be an extra term, namely $  \delta_{ii}=3 $, while no additional term is generated for the traceless transformations:
\begin{align}\label{trafo}
 \left[-3 + \sum_{a=1}^{n} \vec k_{a}\cdot \vec \partial_{a} +f^{LO\delta}(k_{a}) \right] B_{n}=0\,,\\
 \sum_{a=1}^{n}\left[  k^{<i}_{a} \partial_{k^{j>}_{a}} +  f^{LOT}(k_{a})k^{<i}_{a}k^{j>}_{a} \right] B_{n}=0\,.\label{trafo1p5}
\end{align}
We know already that one possibility are dilations, which correspond to $ f^{LO\delta}=3-\Delta$. Requiring that all correlators obey scale invariance, we find
\begin{align}\label{trafo2}
 \left[-3 + \sum_{a=1}^{n} \vec k_{a}\cdot \vec \partial_{a} + 3 - \Delta \right] B_{n}= \left[\left( 3n-\Delta n -3 \right) + \sum_{a=1}^{n} \vec k_{a}\cdot \vec \partial_{a} \right] B_{n}=0\,.
\end{align}
This is solved if each correlator scales as $ B_{n}\sim k^{-3(n-1)+\Delta n}$. Assuming locality, no other degree one symmetries can exist with this trace structure, except those that differ from a dilation by the addition of any translation. Let's see what happens to the symmetric traceless transformations. Since dilations are a symmetry, all terms in the symmetric traceless transformation must have the same scaling dimension, namely the scaling dimension of $ k^{<i}\partial_{j>} $, which is zero. Failure to meet this criterion would imply and infinite set of generators, forbidding any non-vanishing correlators. Therefore also $ f^{LOT}k^{<i}k^{j>} $ must have scaling dimension zero. But if we insist on locality, i.e.~the absence of inverse Laplacians, then $  f^{LOT} $ must vanish. We can now recall that the only power spectrum invariant under rotations, dilations and translations is  $  P=k^{-3+2\Delta} $. It is easy to check that under the transformation in \eqref{trafo1p5} one finds
\begin{align}
\sum_{a}^{2}k_{a}^{<i}\partial_{k_{a}^{j>}} \frac{1}{k_{1}^{3-2\Delta}} =\left(  2\Delta-3 \right)\frac{k_{1}^{<i}k_{1}^{j>}}{k_{1}^{5-2\Delta}} \overset{!}{=} 0\,.
\end{align}
So the power spectrum is invariant only if $  \Delta =3/2 $, in which case $  P(k) $ is a constant and in real space there are no correlations at separated points. We conclude that this cannot be a symmetry of any theory with two-point correlations at separated points and therefore we exclude it from our classification.

 
\subsection{Degree two and higher}

Let's continue to degree-two symmetries, $  S^{2} $. The commutator of any $  S^{2} $ with translations must give the only symmetry of degree one that we found, namely rotations and dilations and so  
\begin{align}\label{comm}
[S^{2}_{i},P_{j}]=a M_{ij}+ b \delta_{ij} D\,,
\end{align}
where $  a $ and $  b $ are some constants and by consistency $  S^{2} $ must have precisely one spatial index
\begin{align}
S^{2}_{i}:\delta \phi= \left[ f^{ijl}\partial_{j}\partial_{l}+f^{ij}\partial_{j}+f^{i} \right]\phi \,.
\end{align} 
Using \eqref{31} we also know that $  S^{2}_{i} $ must have scaling dimension $ -1  $, so that, once commuted with translations, which have scaling dimension 1, it can give dilations or rotations, both with scaling dimension 0. Using again locality this implies $  f^{i}=0 $ and the scalings $  f^{ijl}\sim k^{1} $ and $  f^{ij}\sim k^{0} $. Hence, the most general for of $  S^{2} $ is
\begin{align}
S^{2}:\delta \phi(\k)= \left[ C_{1}\vec{k} \cdot \vec{\partial} \partial_{i}+C_{2}k_{i}\partial^{2}+C_{3} \partial_{i} \right]\phi(\k) \,.
\end{align}
Substituting this form into the commutator \eqref{comm}, gives
\begin{align}
C_{1}&=-2C_{2}=a=-b\,,&C_{3}=-b(3-\Delta)\,.
\end{align}
A convenient rescaling of $  S^{2} $ is $  a=2i $ and then one finds
\begin{align}
S^{2}=K^{i}:\delta \phi(\k)= i\left[ 2 \vec{k} \cdot \vec{\partial} \partial_{i}-k_{i}\partial^{2}+2\left(  3-\Delta\right) \partial_{i} \right]\phi(\k) \,,
\end{align}
which we recognize as a special conformal transformation, satisfying the commutator
\begin{align}
[P_{i},K_{j}]=2D\delta_{ij}+2M_{ij}\,.
\end{align}
This is the only admissible symmetry of degree two.

As proven in \cite{Pajer:2018egx}, no other symmetry of degree three or higher exists that extends the conformal group. The strategy to prove this fact is to show that, if the only degree-two symmetries are special conformal transformation, as in the case at hand, than there is no degree three symmetry that can obey the appropriate commutation relations with translations as well as the Jacobi identities. We can therefore conclude that our classification of symmetries is complete. In summary, for a general scalar field $  \phi $ we found that, assuming invariance under rotations, translation and dilations, the only additional linearly-realized finite-degree symmetries we can add without making all correlators vanish are special conformal transformations. This concludes the proof of Theorem 3.

 
\subsection{Contractions and discrete symmetries}\label{sec:}

Summarizing, we have found that the largest possible algebra of symmetries, given the assumptions in above \eqref{ass}, is given by the conformal group in three euclidean dimensions, a.k.a. the Lorentz group SO$  (3,1) $:
\begin{align}
[M_{ij},M_{kl}]&=4\delta_{[i}^{[k}M_{j]}^{l]}\,, & [M_{ij},P_{l}]&=2\delta_{l[i}P_{j]}\,, & [P_{i},P_{j}]&=0\,, \\
[D,P_{i}]&=P_{i}\,, &  [D,M_{ij}]&=0\,, & [D,K_{i}]&=-K_{i}\,, \\
[K_{i},K_{j}]&=0\,, & [M_{ij},K_{l}]&=2\delta_{l[i}K_{j]}\,, & [P_{i},K_{j}]&=2M_{ij}+2D\delta_{ij}\,.
\end{align}
This algebra admits some well-known Wigner-In\"on\"u contractions, in which some of the commutators are set to zero and a new consistent algebra emerges. The most well-known case is the Galilean algrebra, but all possibilities were classified in \cite{Bacry:1968zf}. It is natural to ask why we did not encounter any of these contractions in our classification. To see this, recall that a contraction consists of keeping all the generators of a given subalgebra fixed and rescaling the others by a parameter that is then taken to zero, so that the rescaled generators become an abelian subalgebra. In doing this we want to require that rotations act in the standard way, so we look for all subalgebras that contain $  M_{ij} $. There are six possibilities, $  \{M_{ij}\} $, $  \{M_{ij},P_{l}\} $, $  \{M_{ij},K_{l}\} $, $  \{M_{ij},D\} $, $  \{M_{ij},P_{l},D\} $ and $  \{M_{ij},K_{l},D\} $. Contractions with respect to subgroups that do not contain $  D $, i.e. involving the rescaling $  D\to D'=\e D $, have the effect to set to zero the commutator with translations $  [D',P_{i}]=0 $. This is inconsistent with the way translations and dilations act on correlators and so this possibility is excluded by assumption. The other three contractions have the effect of setting to zero the commutator $  [P_{i}',K_{j}']=0 $. Then $  K_{i}' $ becomes a degree zero symmetry, which we have proven cannot exist as long as there is any non-vanishing correlators. Hence this discussion confirms our findings in Theorem 3: although there are many other consistent algebras that are obtained by contractions of the conformal algebra, none of them can be realized on non-trivial scalar correlators.

While so far we have only discussed continuous symmetries, it's worth mentioning two discrete symmetries that can be realized on a single scalar $  \phi $, namely space inversion $  P $ and internal reflection $  Q $ acting as
\begin{align}
P&: \phi(\x)\to \pm \phi(-\x)\,, &Q&:\phi(\x)\to -\phi(\x)\,.
\end{align}
The internal discrete symmetry $  Q $ commutes with all other generators, while the spacetime discrete symmetry $  P $ commutes with rotations and dilations, which are even, but anti-commutes with translations and de Sitter boosts, which are odd. Clearly all $  n $-point correlators of $  \phi $ are invariant under $  Q $ if $  n $ is even and so can be non-vanishing. Parity can be broken only if the parity odd but rotational invariant combination $  \left( \vec {k}_{a}\times \vec {k}_{b} \right)\cdot \vec{k}_{c} $ appears.

 
\section{Discussion and conclusions}\label{sec:conc}

Our understanding of the very early universe is informed by the correlations of cosmological perturbations observed at much later times.  It is our hope that the structure of these correlations is sufficiently restrictive that, from them, we can determine the mechanism for inflation and some of the laws of physics at high energies.  In this paper, we explored constraints on the structure of correlators imposed by symmetries and how they relate to the underlying particle content and mechanism for inflation. Given cosmological observations, we find that conformal invariance is the largest space-time symmetry that can act linearly and non-trivially on curvature perturbations.  Furthermore, in single-clock inflation, only scale invariance is possible in an interacting theory, but not linearly-realized special conformal transformations.  

These results contribute to the larger goal of putting our understanding of cosmological correlators on par with boundary correlators in asymptotically flat (S-matrix) or anti-de Sitter space (CFT).  In those cases, the structure of the S-matrix and CFT correlators is sufficiently rigid that it implies a number of non-trivial constraints on physics in the bulk. Holography in de Sitter is a less useful tool, particularly due to the lack of unitarity in dS/CFT~\cite{Witten:2001kn,Strominger:2001pn,Mazur:2001aa,Maldacena:2002vr,Maldacena:2011nz}.  Nevertheless, recent work has used the conformal symmetry of quantum fields in de Sitter~\cite{Arkani-Hamed:2018kmz,Baumann:2019oyu,Sleight:2019mgd,Sleight:2019hfp} and/or the structure of perturbative calculations~\cite{Arkani-Hamed:2015bza,Arkani-Hamed:2018bjr,Benincasa:2018ssx,Benincasa:2019vqr} to draw broader insights in these observables.  The results presented here advance this program further, working directly in terms of late-time $\zeta$ correlations without appealing directly to the local dynamics during inflation or to any Lagrangian description.  Our results imply constraints on dynamics during inflation that are hardly transparent at the level of the action. A priori, $\zeta$ could have Lorentz-invariant couplings to some massive field of any spin that could produce a conformal, non-Gaussian correlation function. Yet, the results presented here show that such a coupling must always break Lorentz / de Sitter invariance, no matter how ingenious the model.

Famously, causality and unitarity for scattering amplitudes~\cite{Adams:2006sv,Camanho:2014apa} or CFT correlators~\cite{Hartman:2015lfa,Afkhami-Jeddi:2016ntf,Cordova:2017zej} place surprisingly strong constraints on coupling constants that are consistent with the symmetries of an EFT.  Most notably, self-consistency can constrain an infinite list of couplings in terms of a single parameter.  Attempts to apply the same techniques to inflation suggest that the couplings of the EFT of inflation are bounded in terms of the speed of propagation of $\zeta$, $c_s$~\cite{Baumann:2015nta,Baumann:2019ghk,Pajer:2020,Grall:2020}.  Most dramatically, it has been conjectured that when $c_s =1$, the theory is necessarily Gaussian up to slow-roll corrections.  This conjecture shares some relation to the present work, as $c_s =1$ corresponds to an enhanced conformal symmetry for the quadratic action.  Our results show that there are no additional operators that can be added to this theory that preserve this symmetry; however, that still leaves the majority of operators that break the symmetry that also give rise to non-Gaussian correlators.  While useful constraints can be derived using limits where scattering or holographic tools apply, we would like to understand the constraints on these EFTs from cosmological correlators alone.  

Although inflationary correlators appear to lack the rigid structure associated with the S-matrix or CFT correlators needed to derive structural constraints, this work has shown that this concern may be overstated.  The single-field consistency conditions are the statement that cosmological correlators obey a nonlinearly realized conformal symmetry, and this symmetry can be put to work to remove some of the ambiguities present for correlation functions of generic fields such as field redefinitions.  Given recent work understanding the analytic structure of these correlators and their connection to the S-matrix~\cite{Maldacena:2011nz,Raju:2012zr}, it seems reasonable to expect a future where model-independent constraints on the dynamics during inflation are derived from cosmological correlators directly.

 
\section*{Acknowledgements}\label{sec:}

We are grateful to Daniel Baumann, Paolo Benincasa, Giovanni Cabass, Paolo Creminelli, Raphael Flauger, Victor Gorbenko, Tanguy Grall, Tom Hartman, Sadra Jazayeri, Scott Melville, Rafael Porto, and David Stefanyszyn for helpful discussions. We would like to thank also the participants of the `Amplitudes meet Cosmology' workshop\footnote{\small \url{ https://www.simonsfoundation.org/event/amplitudes-meet-cosmology-2019/}} for useful conversations. D.\,G.~is supported by the US~Department of Energy under grant no.~DE-SC0019035. E.\,P.~has been supported in part by the research program VIDI with Project No.~680-47-535, which is (partly) financed by the Netherlands Organization for Scientific Research (NWO).


\clearpage
\phantomsection
\addcontentsline{toc}{section}{References}
\bibliographystyle{utphys}
\bibliography{conformal,refs}

\end{document}